\documentclass[aps,prx,twocolumn,preprintnumbers,amsmath,amssymb,superscriptaddress]{revtex4-2}
\usepackage{graphicx}
\usepackage[caption=false]{subfig}
\usepackage{epsfig}
\usepackage{dcolumn}
\usepackage{bm}
\usepackage{ulem}
\usepackage{color}
\usepackage{float}
\usepackage[colorlinks]{hyperref}
\usepackage{verbatim}
\usepackage{algorithm}
\usepackage{algorithmic}
\usepackage{booktabs}

\hypersetup{citecolor = blue}

\def\bra#1{\left\langle#1\right|}
\def\ket#1{\left|#1\right\rangle}

\def\be{\begin{equation}}       \def\ee{\end{equation}}
\def\bea{\begin{eqnarray}}      \def\eea{\end{eqnarray}}
\def\ba{\begin{array}}
\def\ea{\end{array}}
\def\bnum{\begin{enumerate} }
\def\enum{\end{enumerate}}

\def\=>{\Rightarrow}
\def\>{\rightarrow}

\def\eye2{Fathbb{I}}

\def\Eq#1{Eq.~(\ref{#1})}

\renewcommand{\>}{\rangle}

\newcommand{\al}[1]{\begin{align}#1\end{align}}
\newcommand{\eq}[2]{
	\begin{equation}
	#1 \label{#2}
	\end{equation}
}

\newcommand{\re}[1]{\frac{1}{#1}}

\newcommand{\vect}[1]{\boldsymbol{#1}}

\usepackage{listings}
\usepackage{color}
\definecolor{lightgray}{gray}{1}

\begin{document}

\title{Differentiable Quantum Architecture Search}

\author{Shi-Xin Zhang}
\affiliation{Institute for Advanced Study, Tsinghua University, Beijing 100084, China}
\affiliation{Tencent Quantum Laboratory, Tencent, Shenzhen, Guangdong, China, 518057}
\author{Chang-Yu Hsieh}
\email{kimhsieh@tencent.com}
\affiliation{Tencent Quantum Laboratory, Tencent, Shenzhen, Guangdong, China, 518057}
\author{Shengyu Zhang}
\affiliation{Tencent Quantum Laboratory, Tencent, Shenzhen, Guangdong, China, 518057}
\author{Hong Yao}
\email{yaohong@tsinghua.edu.cn}
\affiliation{Institute for Advanced Study, Tsinghua University, Beijing 100084, China}
\affiliation{State Key Laboratory of Low Dimensional Quantum Physics, Tsinghua University, Beijing 100084, China}

\begin{abstract}
Quantum architecture search (QAS) is the process of automating architecture engineering of quantum circuits.
It has been desired to construct a powerful and general QAS platform
which can significantly accelerate current efforts to identify quantum advantages of error-prone and depth-limited quantum circuits in the NISQ era.
Hereby, we propose a general framework of differentiable quantum architecture search (DQAS), which enables automated designs of quantum circuits in an end-to-end differentiable fashion.
We present several examples of circuit design problems to demonstrate the power of DQAS.
For instance, unitary operations are decomposed into quantum gates, noisy circuits are re-designed to improve accuracy, and circuit layouts for quantum approximation optimization algorithm are automatically discovered and upgraded for combinatorial optimization problems.
These results not only manifest the vast potential of DQAS being an essential tool for the NISQ application developments, but also present an interesting research topic from the theoretical perspective as it draws inspirations from the newly emerging interdisciplinary paradigms of differentiable programming, probabilistic programming, and quantum programming.
\end{abstract}

\date{\today}
\maketitle
\renewcommand\thesection{\arabic{section}}

\section{Introduction}

In the noisy intermediate-scale quantum technology (NISQ) era \cite{Preskill2018}, the hybrid quantum-classical (HQC) computational scheme, combining quantum hardware evaluations with classical optimization outer loops, is widely expected to deliver the first instance of quantum advantages (for certain non-trivial applications) in the absence of fault-tolerant quantum error corrections.
Several prototypical examples in this category include finding the ground state of complex quantum systems by variational quantum eigensolver (VQE) \cite{Peruzzo2014, McClean2016,McArdle2020},  exploring better approximation for NP hard combinatorial optimization problems by quantum approximation optimization algorithms (QAOA) \cite{Farhi2014,Hadfield2017,Zhou2020}, and solving some learning tasks in either the classical or quantum context by the quantum machine learning (QML) setup \cite{Farhi2018,Verdon2019, Benedetti2019a,Benedetti2019b,Cong2019,Verdon2019}.

Under the typical setting in the HQC computational paradigm, the structure of variational ansatz is held fixed and only trainable parameters are optimized to satisfy an objective function.
This lack of flexibility is rather undesirable as different families of parametrized circuits may differ substantially in their expressive power and entangling capability \cite{Akshay2020,Farhi2020}.
Moreover, in the NISQ era, a thoughtful circuit design should minimize the consumption of quantum resources due to decoherence and limited connectivity among qubits in current quantum hardwares.
For instance, the number of two-qubits gates (or the circuit depth) should be minimized to reduce noise-induced errors.
Additional error mitigation strategy should be conducted without using extra qubits if possible.
With these requirements in mind, the design of an effective circuit ansatz should take into account of the nature of the computational problems and the specifications of a quantum hardware as well.
We term the automated design of parameterized circuits, in the aforementioned setting,  as quantum {\it ansatz} search (QAS).

In a broader context, we denote QAS as quantum {{\it architecture} search, which covers all scenarios of quantum circuit design and goes beyond the design of a variational ansatz for HQC algorithms.
QAS can facilitate a broad range of tasks in quantum computations.
Its applications include but not limited to decomposing arbitrary unitary \cite{Kiani2020} into given quantum gates, finding possible shortcuts for well-established quantum algorithms \cite{Li2017c, Cincio2018a}, exploring optimal quantum control protocols \cite{Yang2017, Fosel2018, Lin2019a}, searching for powerful and resource-efficient variational ansatz \cite{Rattew2019}, and designing end-to-end and task-specific circuits which also incorporate considerations on quantum error mitigation (QEM), native gate set, and topological connectivity of a specific quantum hardware \cite{Chivilikhin2020, Cincio2020}.

Neural architecture search (NAS) \cite{Yao2018}, devoted to the study and design of neural networks shares many similarities with designing parameterized quantum circuits.
The common approaches for NAS include greedy algorithms \cite{Huang2018}, 
evolutionary or genetic algorithms \cite{Stanley2019, Real2017a, Xie2017, Liu2018a, Real2019a}, and reinforcement learning (RL) based methods \cite{Zoph2017, Baker2017, Cai2018, Zoph2018}. It is interesting to witness that the progress in QAS follows closely the ideas presented in NAS.
Recent works on quantum circuit structure or ansatz design also exploited greedy methods \cite{Ostaszewski2019, Grimsley2019,Li2020}, evolutional or genetic methodologies \cite{LasHeras2016, Li2017c,Cincio2018a, Rattew2019,Cincio2020,Chivilikhin2020} and RL engine based approaches \cite{Fosel2018, Niu2019} for tasks such as quantum control, QEM or circuit ansatz searching.

Recently, differentiable neural architecture search (DARTS) has been proposed \cite{Liu2019a} and further refined with many critical improvements and generalizations \cite{Xie2019,Liang2019, Zela2019,Hundt2019,Chen2019,Casale2019}.
The key idea of a differentiable architecture search is to relax the discrete search space of neural architectures onto a continuous and differentiable domain, rendering much faster end-to-end NAS workflow than previous methods.
Due to the close relation between NAS and QAS, it is natural to ask
whether it is possible to devise a differentiable quantum architecture search (DQAS) incorporating DARTS-like ideas.
Our answer is affirmative;  
as presented in this work, we constructed a general framework of DQAS that works very well as a universal and fully automated design tool for quantum circuits. 
As a general framework sitting at the intersection of newly emerging interdisciplinary paradigms of differentiable programming, probabilistic programming and quantum programming, DQAS is of both high theoretical and practical values across various fields in quantum computing and quantum information processing.

The organization of this work goes as follows. In Background and Related Work section, we review backgrounds and relevant works on fields including NAS, QAS and QAOA. In Methods section, we introduce the setup of the DQAS algorithm, where the overall workflow and the main components are both discussed. In Applications section, we demonstrate various applications in quantum computing domain enabled by DQAS, including QEM and variational quantum algorithm design examples. We conclude with a brief Discussion section. The Appendix contains more details and further applications of DQAS \footnote{See Appendix for more information on:  A. Glossary for ingredients of DQAS. B. The connection betweeen DARTS and DQAS.  C. Gradients derivation from Monte Carlo expectation in DQAS. D. General hyperparameters for DQAS training E. Training techniques that stabilize and improve DQAS. F. DQAS applications on state preparation and unitary learning. G. Relevant hyperparameter and ingredient settings on experiments in this work. H. More results and comparisons on QEM for QFT circuit.  I. More DQAS  results for QAOA ansatz search including instance learning, block encoding, etc.}.

\section{Background and Related Work} \label{sec:background}
\noindent{\bf Differentiable Neural Architecture Search.}
NAS \cite{Ren2020} is a burgeoning and active field in AutoML, and the ultimate goal of NAS is to automate the search for a top-performing neural network architectures for any given task.
Popular approaches to implement NAS include reinforcement learning \cite{Zoph2017}, in which an RNN controller chooses an action on building the network structure layerwise from a discrete set of options;
and evolutionary or genetic algorithms \cite{Real2017a, Xie2017,Real2019a}, in which a population of network architectures is kept, evaluated, mutated for the fittest candidates.
Such RL or evolutionary algorithms are rather resource intensive and time consuming,
since the core task involves searching through an exponentially large space of discrete choices for different elementary network components.

Recently, differentiable architecture search \cite{Liu2019a} and its variants have been proposed and witnessed a surge in the number of related NAS studies \cite{Xie2019,Liang2019, Zela2019,Hundt2019,Chen2019,Casale2019, Dong2019,Yao2019,Noy2019,Cai2019, Li2019,Xu2019, Chang2019,Chen2020, Hu2020}.
Under the DARTS framework, the network architecture space of discrete components is relaxed into a continuous domain that facilitates search by differentiation and gradient descent.
The relaxed searching problem can be efficiently solved with noticeably reduced training time and hardware requirements.

In the original DARTS, the search space concerns with choices of distinct microstructures within one cell. Two types of cell are assumed for the networks: normal cell and reduction cell.
The NAS  proceeds by first determining the microstructures within these two types of cell, then a large network is built by stacking these two cell types up to a variable depth with arbitrary input and output size.
Within each cell, two inputs, four intermediate nodes and one output (concatenation of four intermediate nodes) are presented as nodes in a directional acylic graph.
For each edge between nodes, one needs to determine optimal connection layers, eg. conv with certain kernel size, or max/average pooling with given window size, zero/identity connections and so on.
To make such search process differentiable, we assume each edge is actually the weighted sum of all these primitive operations from the pool, i.e. $o(x) = \sum_i \text{softmax}(\alpha_i) o_i(x)$ where $o_i$ stands for  i-th type of layers primitives and $\alpha_i$ is the continuous weights which determines the structure of neural network as structural parameters.
Therefore, we have two sets of continuous parameters: structure weights $\alpha$ which determines the optimal network architecture by pruning in evaluation stage, and conventional parameters in neural network $ \omega$.
Via DARTS setup, neural architecture search turns into a bi-optimization problem where differentiation is carried out end-to-end.

DARTS requires the evaluation of the whole super network where each edge is composed of all types layers.
This is memory intensive and limits its usage on large dataset or enriched cell structures.
Therefore, there are works extending DARTS idea while enabling forward evaluation on sub network, usually using only one path \cite{Casale2019,Hu2020} or two \cite{Cai2019}.
Specifically, in \cite{Casale2019}, the authors viewed the super network as a probabilistic ensemble of subnetworks and thus variational structural parameters enter into NAS as probabilistic model parameters instead.
So we can sample subnetworks from such probabilistic distribution and evaluate one subnetwork each time.
This is feasible as probabilistic model parameters can also be updated from general theory for Monte Carlo expectations' gradient in a differentiable approach \cite{Mohamed2019a}.

There are additional follow-up works that focus on improving drawbacks of DARTS with various training techniques.
In general, these DARTS-related techniques are also illuminating and inspirational for further DQAS developments in our work.

~\newline	
\noindent{\bf Related works on QAS.} Quantum architecture search, though no one brand it as this name before, is scattered in the literature with different contexts.
These works are often specific to problem setup and denoted as quantum circuit structure learning \cite{Ostaszewski2019}, adaptive variational algorithm \cite{Grimsley2019},  ansatz architecture search \cite{Li2020}, evolutional VQE \cite{Rattew2019}, multipleobjective genetic VQE \cite{Chivilikhin2020} or noise-aware circuit learning \cite{Cincio2020}.
The tasks they focused are mainly in QAOA \cite{Li2020} or VQE \cite{Ostaszewski2019, Grimsley2019, Rattew2019, Chivilikhin2020} settings.
From higher theoretical perspective, some quantum control works can also be classified as QAS tasks, where optimal quantum control protocol is explored using possible machine learning tools \cite{Fosel2018, Niu2019} .

These QAS relevant works are closed related to NAS methodologies.
And this relevance is as expected, since quantum circuit and neural network structure share a great proportion of similarities.
The mainstream approach of QAS is evolution/genetic algorithms with different variants on mutation, crossover or tournament details \cite{LasHeras2016, Li2017c,Cincio2018a, Rattew2019,Cincio2020,Chivilikhin2020}.
There are also works exploiting simple greedy/locality ideas \cite{Ostaszewski2019, Grimsley2019,Li2020} and reinforce learning ideas \cite{Fosel2018, Niu2019}.

All of the QAS works mentioned above are still searching quantum ansatz/architecture in discrete domain, which increases the difficulty on search and is in general time consuming.
Due to the close relation between QAS and NAS together with the great success of differentiable NAS ideas in machine learning, we here introduce a framework of differentiable QAS that enable end-to-end automatic differentiable QAS (DQAS).
This new approach unlocks more possibilities than previous works with less searching time and more versatile capabilities.
It is designed with general QAS philosophy in mind, and DQAS framework is hence universal for all types of circuit searching tasks, instead of focusing only on one type of quantum computing tasks as previous work.

~\newline
\noindent{\bf Brief review on QAOA.} 
As introduced in \cite{Farhi2014}, QAOA is designed to solve classical combinatorial optimization (CO) problems.
These problems are often NP complete, such as MAX CUT or MIS in the graph theory.
The basic idea is that we prepare a variational quantum circuit by alternately applying two distinct Hamiltonian evolution blocks.
Namely, a standard QAOA anstaz reads
\eq{
	\vert \psi\rangle =\prod_{j=0}^{P} (e^{iH_c \gamma_j}e^{iH_b\beta_j})\vert \psi_0\rangle,
}{vanilla-qaoa}
where $\vert\psi_0\rangle$ should be prepared in the space of feasible solutions (better as even superposition of all possible states, and in MAX CUT case $\vert\psi_0\rangle = \otimes H^n\vert 0^n\rangle$, where n is the number of qubits and H is transversal Hadamard gates.)

In general, $H_c$ is the objective Hamiltonian as $H_c\vert\psi\rangle = f(\psi)\vert\psi\rangle$, where $f(\psi)$ is the CO objective. For MAX CUT problem on weighted graph with weight $\omega_{ij}$ on edge $ij$, $H_c=-\sum \omega_{ij}Z_iZ_j$ up to some unimportant phase.
(We use the notation $X/Y/Z_i$ for Pauli operators on i-th qubit throughout this work)
$H_b$ is the mixer Hamiltonian to tunnel different feasible solutions, where $H_b=\sum_{i=0}^n X_i$ is the most common one when there is no limitation on feasible Hilbert space.

The correctness of such ansatz is guaranteed when p approach infinity as it can be viewed as quantum annealing (QA), where we start from the ground state of Hamiltonian $H_b$ as $\vert +^n\rangle$ and go through adiabatically to another Hamiltonian $H_c$, then it is expected that the final output state is the ground state of $H_c$ which of course has the minimum energy/objective and thus solve the corresponding CO problems.

If we relax the strong restrictions from the QA limit and just take QAOA as some form of variational ansatz, then there are four Hamiltonians instead of two defining the ansatz.

\begin{itemize}
	\item $H_p$ the preparation Hamiltonian: we should prepare the initial states from zero product to the ground state of $H_p$. In original case, $H_p$ is the same as $H_b$.
	\item $H_b$ the mixer Hamiltonian: responsible to make feasible states transitions happen.
	\item $H_p$ the phase/problem Hamiltonian: time evolution under the phase Hamiltonian and the mixer Hamiltonian alternately makes the bulk of the circuit, in original QAOA, $H_p$ is the same as $H_c$.
	\item $H_c$ the cost Hamiltonian: the Hamiltonian used in objectives and measurements where $\langle \psi \vert H_c\vert \psi\rangle$ is optimized.
\end{itemize}

Moreover, such four Hamiltonian generalization of original QAOA can be further extended.
For example, $H_b, H_p$ are not necessarily the same Hamiltonian for each layer of the circuit.
Nonetheless, the essence of such ansatz is that: the number of variational parameters is of order  the same as layer number $P$ which is much less than other variational ansatz of the same depth such as typical hardware efficient VQE or quantum neural network design.
This fact renders QAOA easier to train than VQE of the same depth and suffers less from barren plateus \cite{McClean2018}.
And as QAOA ansatz has some reminiscent from QA, the final ansatz has better interpretation ability than typical random circuit ansatz.
It is an interesting direction to automatically search for the four definition Hamiltonians or even more general layouts beyond vanilla QAOA, to see whether there are similar quantum architectures that can outperform vanilla QAOA in CO problems, this is where DQAS plays a role.

The physical intuition behind such QAOA type ansatz relaxation and searching originates from the close relation between QAOA and quantum adiabatic annealing.
In particular, we draw inspirations from efforts to optimize annealing paths and boost performance for quantum annealers.
We reckon at least two fronts to search for better ansatz for the hybrid quantum-classical algorithm.  The first case is to actually inspect the standard QAOA (which typically uses two alternating Hamiltonians to build the ansatz) and inquire if any ingredient may be improved.  For instance, given the four Hamiltonians for the quantum-adiabatic inspired ansatz, one may search for a better initial-state-preparation Hamiltonian, or find better mixer Hamiltonians than the plain $\sum_i X_i$ for specific problems.
Another inspiration derives from attempts to speed up quantum adiabatic annealing via ideas like catalyst Hamiltonians \cite{Crosson2014, Albash2019}, counter diabatic Hamiltonians \cite{Sels2017, Hartmann2019}, and other ideas in shortcut to adiabacity.
These ultrafst annealing methods would entail design of complex annealing schedules that deviate from the simple linear schedule interpolating between an initial Hamiltonian and the target Hamiltonian.
When these complex annealing paths are digitalized and projected onto the quantum gate model with variational approximations,  they may just live in the form of XX Hamiltonians or local Y Hamiltonians. With these extra Hamiltonians, catalyst or counter diabatic, we anticipate better performances with shallower QAOA-like circuits layout may be achieved.

\section{Methods}

\noindent{\bf Overview.}  The task of DQAS is to select several unitaries to compose the circuit that minimize the corresponding objective for a given task. The aim of DQAS is two-fold: one the one hand, DQAS determines a potentially optimal circuit layout, on the other hand, it also identifies suitable variational parameters for the circuit. To achieve the two goals at the same time, we regard DQAS as a bi-optimization problem, where both the parameters determining the quantum structure and trainable weights on the parameterized circuit are optimized via some gradient-based optimizers. 
To enable gradient descent search on the quantum structure, we relax the discrete structure parameters into continous domain, where quantum architecture are viewed as the sample from some parameterized probabilistic model.

DQAS is presented as Algorithm \ref{alg:dqas} with a visualized workflow in Fig. \ref{fig:workflow}. We introduce the ingredients for DQAS and the general workflow below. (See Appendix A for more details and the glossary of DQAS algorithm \cite{Note1}) .

\begin{figure}[t]
	\includegraphics[width=0.47\textwidth]{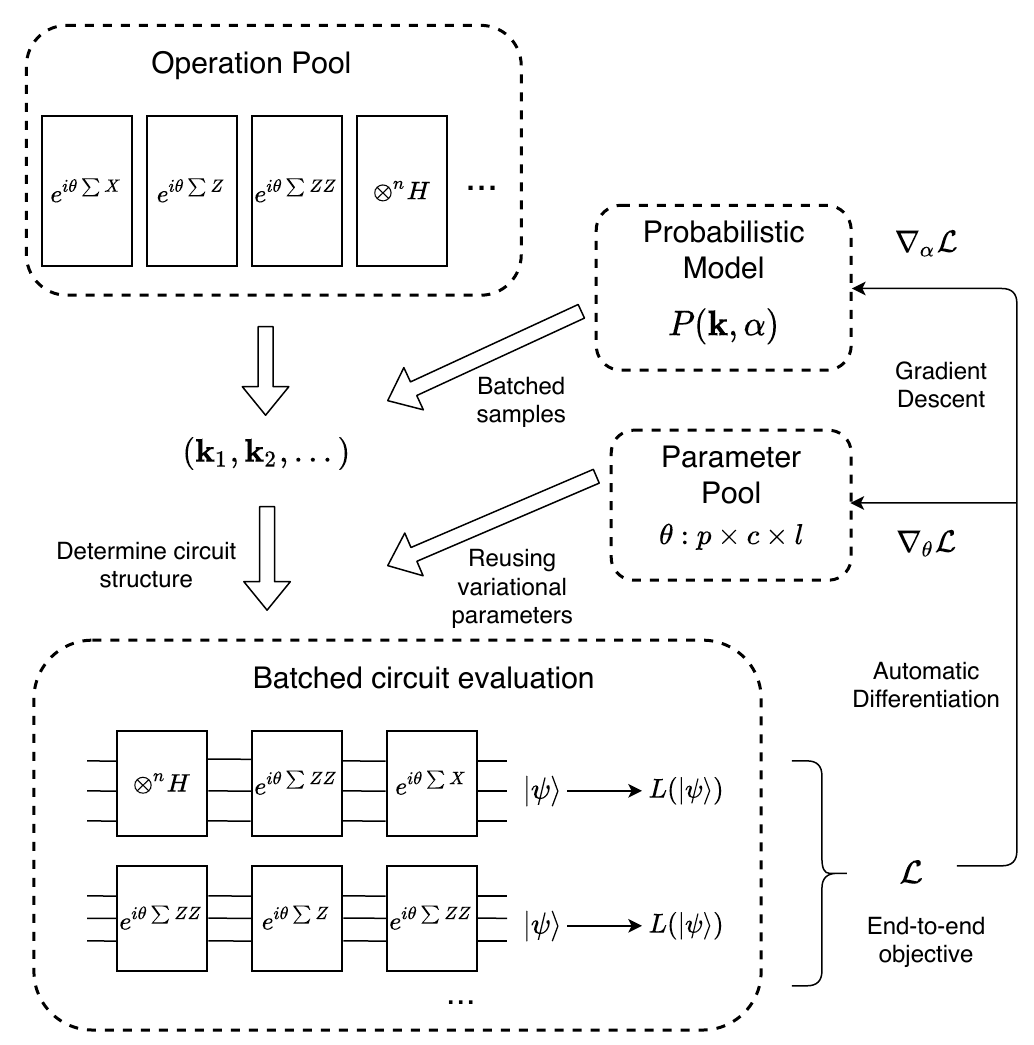}
	\caption{Schematic illustration of DQAS.
		We sample a batch of circuit configurations for each epoch from probabilistic model $P(\vect{k}, \vect{\alpha})$.
		We then compose corresponding quantum circuits by filling in operations and parameters from two pools.
		We can evaluate quantum circuits and compute final objective $\mathcal{L}$ where $\vect{\alpha}, \vect{\theta}$ can be adjusted accordingly with gradient based optimization method.
	}\label{fig:workflow}
\end{figure}

~\newline
\noindent{\bf Circuit encoding and operation pool.}  Any quantum circuit is composed of a sequence of unitaries with and without trainable parameters, i.e. 
\eq{U=\prod_{i=0}^p U_i(\vect{\theta}_i), }{circuit}
where $\vect{\theta}_i$ can be of zero length corresponding to the case that $U_i$ is a fixed unitary gate.  Hence, we formulate the framework to cover circuit-design tasks beyond searching of variational ansatz.

In the most general term, these $U_i$ can represent an one-qubit gate, a two-qubit gate or a higher-level block encoding, such as $e^{iH\theta}$ with a pre-defined Hermitian Hamiltonian $H$.
This set of possible unitaries $U_i$ constitutes the operation pool for DQAS, and the algorithm attempts
to assemble a quantum circuit by stacking $U_i$ together in order to optimize a task-dependent objective function.

We call the choice of primitive unitary gates in the operation pool along with the circuit layout of these gates a circuit encoding.
In the operation pool, there are $c$ different unitaries $V_j$, and each placeholder $U_i$ should be assigned one of these $V_j$ by DQAS.
In this work, we refer to the placeholder $U_i$ as the $i$-th layer of the circuit $U$, no matter such placeholder actually stands for layers or other positional labels.
The circuit design comes with replacement: one $V_j$ from the operation pool can be used multiple times in building a single circuit $U$.

~\newline
\noindent{\bf Objectives.} To enable an end-to-end circuit design, a suitable objective should be specified.
Such objectives are typically just sum of expectation values of some observables for hybrid quantum-classical scenarios, such as combinatorial optimization problems or quantum simulations.
Namely, the objective in these cases reads
\eq{L=\langle 0\vert U^\dagger H U\ket{0},}{objective}
where $H$ is some Pauli strings such as $H=-\sum_{\langle ij\rangle} Z_iZ_j$ for MAX CUT problems and $\ket{0}$ represents the direct product state . This loss function $L$ can be easily estimated by performing multiple shots of sample measurements in quantum hardwares.

However, the objectives can assume more general forms for a HQC algorithm. For instance, one may define more sophisticated objectives that not only depend on the mean value of measurements but also depend on distributions of different measurements.
Examples include CVaR \cite{Barkoutsos2020} and Gibbs objective \cite{Li2020}, proposed to improve quality of solutions in QAOA.  In general, DQAS-compatible objectives for HQC algorithms assume the following form, 
\eq{L=\sum_i g_i(\bra{0} U^\dagger f_i(H_i) U\ket{0}),}{gen-fg}
where $f_i$ and $g_i$ are differentiable functions and $H_i$ are Hermitian observables.

Extending the HQC algorithm to supervised machine learning setup that is commonly used in classification tasks, the objective function has to be further generalized to incorporate quantum-encoded data $\ket{\psi_j}$ with corresponding label $y_j$, 
\eq{L=\sum_j\left (\sum_i g_i(\bra{\psi_j} U^\dagger f_i(H_i) U\ket{\psi_j})- y_j\right )^2.}{obj-gen}

Beyond ansatz searching for HQC algorithms, DQAS can be used to design circuits for state preparation or circuit compilations.
In these scenarios, the objective is often taken as the fidelity between the proposed circuit design and a reference circuit, and the objective for pure states now reads 
\eq{L=\sum_j\bra{\phi_{j} }U\ket{\psi_j},}{obj-st}
where $\ket{\phi_j} = U_{\text{ref}}\ket{\psi_j}$ is the expected output of a reference circuit when $\ket{\psi_j}$ is the input state.
For a state-preparation setup, the objective above is reduced to $L=\bra{\phi} U\ket{0}$, where $\ket{\phi}$ is the target state. It is worth noting that the overlap objective can induce barren plateau issues and the local version of Hilbert-Schmidt Test can be used as objectives to avoid barren plateaus \cite{Khatri2019, Sharma2020}.
For a general task of unitary learning or compilation, the dimension of $\ket{\psi_j}$ can be as large as $2^n$ where $n$ is the qubits number, such condition may be relaxed by sampling random inputs from Haar measure \cite{Nakata2017}, which follows the philosophy of machine learning, especially stochastic batched gradient descent.

~\newline
\noindent{\bf Sampling the structures.}  With circuit encoding and operation pool, the task of DQAS is reduced to assign $p$ unitaries (selected from the operation pool) to the placeholder $U_i$ in order to construct a circuit $U$ that minimizes an objective $L(U)$.
To facilitate the architecture search, it is tempting to relax the combinatorial problem into a continuous domain, amenable to optimization via gradient descent.
We thus propose to embed the discrete structural choices into a continuously-parameterized probabilistic model.
For instance, we consider a probabilistic model $P(\vect{k}, \vect{\alpha})$ where $\vect{k}$ is the discrete structural parameter determining the quantum circuits structure and hence $\vect{k}$ is often denoted as an intermediate representation (IR) for quantum circuit structure.
For example, if IR $\vect{k}=[1,3,1]$ then it implies that the circuit structure $U(\vect{k}) = V_1 V_3 V_1$ where $V_1$ and $V_3$ refer to elements in the predefined operation pool introduced earlier. In the context of Eq. \eqref{circuit},  $U_1=V_1$, $U_2=V_3$ and $U_3=V_1$.
$\vect{\alpha}$ is the continuous variable characterizing the distribution of the probabilistic model $P$. 
For na\"ive mean field probabilistic model, $\alpha_{ij}$ stands for the logarithmic probability to place $V_j$ operator on the position of $U_i$ placeholder.
By such a design, we replace the intimidating task of searching for optimal structure in discrete IR space $\vect{k}$ with the easier task of optimizing continuous model parameters $\vect{\alpha}$.

In short, discrete random variables $\vect{k}$ are sampled from a probabilistic model characterized by  parameters $\vect{\alpha}$.
A particular $\vect{k}$ determines the structure of the circuit $U(\vect{k})$, and this circuit is used to evaluate the objectives $L(U)$.  The final end-to-end objectives for DQAS reads 
\eq{\mathcal{L}=\sum_{\vect{k}\sim P(\vect{k}, \vect{\alpha})} L(U(\vect{k}, \vect{\theta})).}{arg2}
And $\mathcal{L}$ depends indirectly on both variational circuit parameters $\vect{\theta}$ and probabilistic model parameters $\vect{\alpha}$, which can be both trained via gradient descent using automatic differentiation.

\begin{algorithm}[H]
	\caption{\small Differentiable Quantum Architecture Search.}
	
	\begin{algorithmic}[1]
		\REQUIRE $p$ as the number of components to build the circuit; operation pool with $c$ distinct unitaries; probabilistic model and its parameters $\vect{\alpha}$ with shape $p\times c$ initialized all to zero; resuing parameter pool $\vect{\theta}$ initialized with given initializer with shape $p\times c\times l$, where $l$ is the max number of parameters of each op in operation pool.
		
		\WHILE{not converged}
		
		\STATE 	~~~~Sample a batch of size K circuits from model $P(k, \vect{\alpha})$.
		
		\STATE ~~~~Compute the objective function for each circuit in the batch in the form of Eq. \eqref{objective}, Eq. \eqref{obj-gen}, Eq. \eqref{obj-st} depending on different  problem settings.
		
		\STATE ~~~~Compute the gradient with respect to $\vect{\theta}$ and $\vect{\alpha}$ according to Eq. \eqref{nablatheta} and Eq. \eqref{nablaalpha}, respectively.
		
		\STATE ~~~~ Update $\vect{\theta}$ and $\vect{\alpha}$ using given gradient based optimizers and learning schedules.

		\ENDWHILE
		
		\STATE Get the circuit architecture $\vect{k}^*$ with the highest probability in $P(\vect{k}, \vect{\alpha})$; and fine tuning the circuit parameters as $\vect{\theta}^*$ associated with this circuit if necessary.
		
		\RETURN final optimal circuit structure labeled by $\vect{k}^*$ and the associating weights $\vect{\theta}^*$.
	\end{algorithmic}
	\label{alg:dqas}
\end{algorithm}

~\newline
\noindent{\bf Filling the circuit parameters.} Since DQAS needs to sample multiple circuits $U$ before deciding whether the current probabilistic model is ideal, we adopt the circuit parameter sharing mechanism for parametrized operators in the operation pool.
We store a tensor of parameters $\vect{\theta}$ with size $p\times c\times l$, where $p$ is the total number/layer of unitary placeholders to build the circuit, $c$ is the size of the operation pool and $l$ is the largest number of parameters for each unitaries in the operation pool, we denoted this tensor as a circuit parameter pool.

For example, if we place the $j$-th operator $V_j$ on the position of placeholder $U_i$ as defined in Eq. \eqref{circuit}, then we should fill such parameterized operator of $l$ parameters with $l$ values from parameter pool: $\theta[i, j, :]$.
Therefore, every sampled parametrized $V_j$ should be initialized with $l$ parameters taken from the circuit parameter pool depending on the placeholder index $i$ and its operation-pool index $j$.  With this circuit parameter sharing mechanism, the variational parameters we need to maintain in architecture search is reduced from $lc^p$ to $lcp$, i.e. an exponential reduction of trainable weights in total. This is the key to enabling a large scale quantum architecture search in terms of the operation pool size and the depth of the circuit. The number of possible quantum architectures is still exponential as $c^p$. However, this exponential scaling in terms of operation pool size is not a severe issue as: (1) the operation pool can be highly customizable and small enough by considering high-level encodings and (2) the exponential space can still be efficiently reached via Monte Carlo sampling from a informed probabilistic model. Therefore, the introduction of parameter sharing and architecture sampling render DQAS as a highly scalable approach for architecture search with moderate resources.

~\newline
\noindent{\bf Quantum and Monte Carlo gradients.} DQAS needs to optimize two sets of parameters, $\vect{\alpha}$ and $\vect{\theta}$,  in order to identify a potentially ideal circuit for the task at hand.
The gradients with respect to trainable circuit parameters $\vect{\theta}$ are easy to determine 
\eq{\nabla_{\vect{\theta}} \mathcal{L}= \sum_{\vect{k}\sim P(\vect{k}, \vect{\alpha})}\nabla_{\vect{\theta}} L(U(\vect{k}, \vect{\theta})).}{nablatheta}
$\nabla_{\vect{\theta}} L(U)$ can be obtained with automatic differentiation in a classical simulation and from parameter shift \cite{Crooks2019} or other analytical gradient measurements \cite{Harrow2019} in quantum experiments.

As explained in Algorithm \ref{alg:dqas},  not all $\theta$ parameters would be present in a circuit
which are sampled according to the probability $P(\vect{k},\vect{\alpha})$ at every iteration.  For missing parameters in a particular circuit,  the gradients are simply set to $0$ as anticipated.

Calculations of gradients for $\vect{\alpha}$ should be treated more carefully, since these parameters are directly related to the outcomes of the Monte Carlo sampling from $P(\vect{k},\vect{\alpha})$.  The calculation of gradient for the Monte Carlo expectations is an extensively studied problem \cite{Mohamed2019a} with two possible mainstream solutions: score function estimator \cite{Kleijnen1996} (also denoted as REINFORCE \cite{Williams1992}) and pathwise estimator (also denoted as reparametrization trick \cite{Kingma2013}).
In this work, we utilize the score function approach as it is more general and bears the potential to support calculations of higher order derivatives if desired \cite{Foerster2018b, Zhang2019b}.
For unnormalized probabilistic model, the gradient with respect to $\vect{\alpha}$ is given by \cite{Note1}
\al{\nabla_{\vect{\alpha}} \mathcal{L}=\sum_{\vect{k}\sim P}\nabla_{\vect{\alpha}} \ln P(\vect{k}, \vect{\alpha})\, L(U(\vect{k}, \vect{\theta}))- \nonumber \\
	\sum_{\vect{k}\sim P}L(U(\vect{k}, \vect{\theta})) \sum_{\vect{k}\sim P}\nabla_{\vect{\alpha}} \ln P(\vect{k}, \vect{\alpha}).\label{nablaalpha}}

For normalized probability distributions, $\langle \nabla_{\vect{\alpha}} \ln P\rangle_P = 0$ and we may simply focus on the first term.
Gradient of $\ln P$ can be easily evaluated via backward propagations on the given well-defined probabilistic model. By considering baseline trick to reduce the estimation variance, a batch size in the order of $10$ is enough for a success DQAS training.

~\newline
\noindent{\bf Probabilistic models.} Throughout this work, we utilize the simplest probabilistic models: independent category probabilistic model, also known as na\"ive mean field model in energy model context.
We stress that more complicated models such as the energy based models \cite{Hinton2012, Carleo2017, Verdon2019} and autoregressive models \cite{Germain2015, Wu2019,Sharir2019, Liu2019} may yield better performances under certain settings where explicit correlation between circuit layers is important.
Such sophisticated probabilistic models can be easily incorporated into DQAS, and we leave this investigation as a future work.

The independent categorical probabilistic model we utilized is described as:
\eq{P(\vect{k}, \vect{\alpha}) = \prod_{i=1}^{p} p(k_i, \vect{\alpha_i}),}{arg2}
where the probability $p$ in each layer is given by a softmax
\eq{p(k_i=j, \vect{\alpha_i)}=\frac{e^{\alpha_{ij}}}{\sum_{k} e^{\alpha_{ik}}},}{arg2}
where $k_i=j$ means that we pick $U_i=V_j$ from the operation pool, and the parameters $\vect{\alpha}$ are of the dimension $p\times c$.
The gradient of such a probabilistic model can be determined analytically,
\eq{\nabla_{\alpha_{ij}} \ln P(k_i=m) = -P(k_i=m) + \delta_{jm}.}{arg2}

\section{Applications}

\noindent DQAS is a versatile tool for near-term quantum computations.  In the following, we present several concrete examples to illustrate DQAS's potential to accelerate research and development of quantum algorithms and circuit compilations in the NISQ era \footnote{Code implementation of DQAS and its applications can be found at \url{https://github.com/refraction-ray/tensorcircuit/tree/master/tensorcircuit/applications}}. Our implementation are based on quantum simulation backend of either Cirq \footnote{See \url{https://github.com/quantumlib/Cirq}}/TensorFlow Quantum \footnote{See \url{https://github.com/tensorflow/quantum}} stack or TensorNetwork \footnote{See \url{https://github.com/google/tensornetwork}}/TensorCircuit \footnote{See \url{https://github.com/refraction-ray/tensorcircuit}} stack.

Firstly, it is natural to apply DQAS to quantum circuits design for state preparation as well as unitary decomposition. For example we can use DQAS to construct exact quantum circuit for GHZ state preparation or Bell circuit unitary decomposition \cite{Note1}. We focus on QEM and HQC context in details below to demonstrate the power of DQAS for NISQ-relevant tasks.

~\newline
\noindent{\bf Quantum error mitigation on QFT circuit.}
Next, we demonstrate that DQAS can also be applied to design noise resilient circuits that mitigate quantum errors during a computation.
The strategy we adopt in this work is to insert single qubit gates (usually Pauli gates) into the empty slots in a quantum circuit, where the given qubit are supposed to be found in idle/waiting status. Such gate-inserting technique can mitigate quantum errors since these extra unitaries (collectively act as an identity operation) can turn coherent errors into stochastic Pauli errors, which are easier to handle and effectively reduce the final infidelity. Similar QEM tricks are reported in related studies \cite{Wallman2016, Zlokapa2020}.

The testbed is the standard circuit for quantum Fourier transformation (QFT), as shown in Fig.~\ref{fig:qem}\subref{fig:qft3}.
We assume the following error model for an underlying quantum hardware.  In between two quantum gates,  there is  a $2\%$ chance of bit flip error incurred on a qubit.  When a qubit is in an idle state (with much longer waiting time), there is a much higher chance of about $20\%$ for bit flip errors.
Although the error model is ad-hoc, it does not prevent us from demonstrating how DQAS can automatically design noise-resilient circuits.

Looking at Fig.~\ref{fig:qem}\subref{fig:qft3}, there are six empty slots in the standard QFT-3 circuit.  Hence, we specify these slots as $p=6$ placeholders for a search of noise-resilient circuits with DQAS.  The search ends when DQAS fills each placeholder with a discrete single-qubit gate such that the fidelity of the circuit's output (with respect to the expected outcome) is maximized in the presence of noises.

If the operation pool is limited to Pauli gates and identity, $\{I, X, Y, Z\}$, then DQAS recommends a rather trivial circuit for error mitigation.
In short, DQAS fills the pair gaps (of qubit 0 and qubit 2) with the same Pauli gate twice, which together yields an identity, in order  to reduce the error in the gap.  As for qubit 1, where a single gap occurs at the beginning and the end of the circuit as shown in Fig.~\ref{fig:qem}\subref{fig:qft3}, DQAS simply fills these gaps with nothing (identity placeholder).  However, if we allow more variety of gates in the operation pool, such as S gate and T gate, then more interesting circuits can be found by DQAS.  For instance, Fig.~\ref{fig:qem}\subref{fig:qft3qem} is one such example.  In this case, DQAS fills the two gaps of qubit 1 with a $T$ gate each. This circuit cannot be found by the simpler strategy of inserting unitaries into consecutive gaps. Thus, DQAS provides a systematic and straightforward approach to identify this kind of long-range correlated gate assignments that  should effectively reduce detrimental effects of noise.

We also carried out DQAS on QFT circuit for 4 qubits with $p=12$ circuit gaps as shown in Fig.~\ref{fig:qft4}\subref{fig:textbookqft4}. 
DQAS automatically finds better QEM architecture which outperforms na\"ive gate inserting policies again. Fig.~\ref{fig:qft4}\subref{fig:qemqft4} displays one such example.
The interesting patterns of long-range correlated gate insertions are obvious for quibt 2.  It is also clear that DQAS learns that more than two consecutive gates can combined collectively to render identity such as the three inserted gates for qubit 0.
Further details on the search for optimal QEM architectures and comprehensive comparison on experiment values of final fidelities can be found in the Appendix H \cite{Note1}.

In summary, DQAS not only learns about inserting pairs of gates as identity into the circuit to mitigate quantum error, but also picks up the technique of the long-range correlated gate assignment to further reduce the noise effects.
This result is encouraging and shows how instrumental DQAS as a tool may be used for designing noise-resilient circuits with moderate consumption of computational resource.
In this study, we only adapt the simple gate-insertion policy to design QEM within DQAS framework.  We expect more sophisticated QEM methods  may also be adapted to work along with DQAS to identify novel types of noise-resilient quantum circuits.  This is a direction that we will actively explore in follow-up studies.

\begin{figure}[t]

	\subfloat[]{\label{fig:qft3}
		\includegraphics[width=0.39\textwidth]{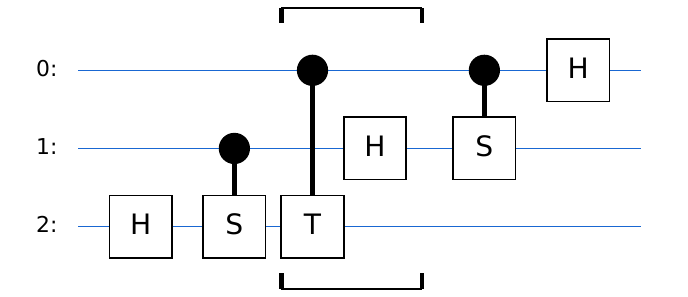}}
	
	\subfloat[]{\label{fig:qft3qem}
		\includegraphics[width=0.39\textwidth]{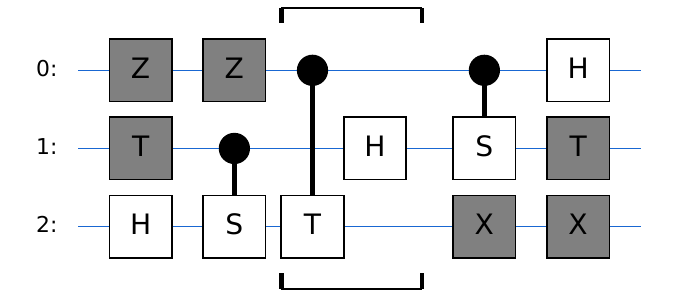}}
	
	\caption{(a) The basic circuit for QFT on 3 qubits, T gate and H gate in the middle of the circuit can be easily arranged in the same vertical moment with no gap. And there are six gaps left (two on each qubit) in this setup. (b) The QEM circuit for QFT automatically found by DQAS. All slots are filled, DQAS is powerful enough to learn long range correlations so that it can fill the gaps on qubit 1 which are seperately located. The fidelity between the two circuits on noisy hardwares and the ideal circuit are  $0.33$ and $0.6$, respectively.
	}\label{fig:qem}
\end{figure}

\begin{figure}[t]

	\subfloat[]{\label{fig:textbookqft4}
		\includegraphics[width=0.43\textwidth]{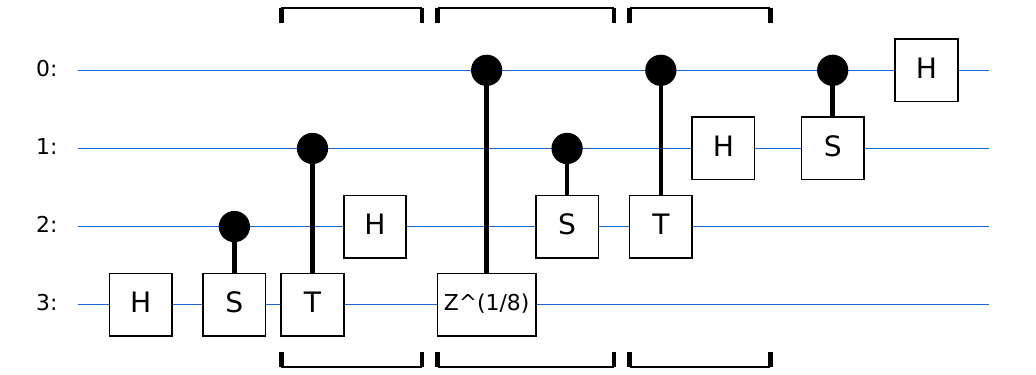}}
	
	\subfloat[]{\label{fig:qemqft4}
		\includegraphics[width=0.46\textwidth]{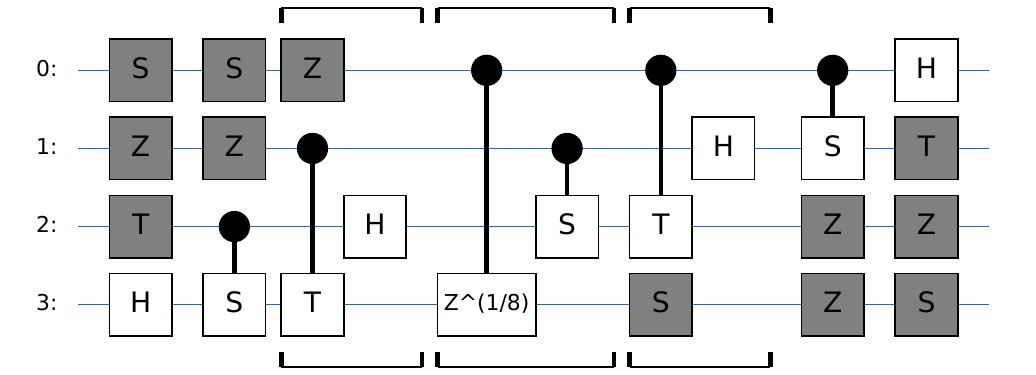}}
	
	\caption{(a) The basic circuit for QFT on 4 qubits, some of the gates can be easily arranged in the same vertical moment with no gap. And there are $12$ gaps left  in this arrangement. (b) The QEM circuit for QFT automatically found by DQAS.  The fidelity between the two circuits on noisy hardwares and the ideal circuit are  $0.13$ and $0.46$, respectively.
	}\label{fig:qft4}
\end{figure}

~\newline
\noindent{\bf QAOA ansatz searching.} QAOA introduces the adiabatic-process inspired ansatz that stacks alternating Hamiltonian evolution blocks as $e^{-i\theta H}$, where $H$ could be different Hermitian Hamiltonians. QAOA can obtain better approximation ratio with increasing number of repetitive circuit blocks $P$ as its infinite $P$ limit is equivalent to quantum adiabatic evolution. To the end of employing DQAS to design parametrized quantum circuits within the hybrid quantum-classical paradigm for algorithmic developments, we adopt a higher-level circuit encoding scheme as inspired by QAOA. More specifically, the operation pool consists of $e^{-i\theta H}$ blocks with different Hermitian Hamiltonians and also parameter free layers of traversal Hadamard gates $\otimes^n H$.  In comparison to assembling a circuit by specifying individual quantum gates, this circuit encoding scheme allows a compact and efficient description of large-scale and deep circuits. For simplicity, we dub the circuit-encoding scheme above as the layer encoding.

For illustrations, we apply DQAS to design parametrized circuit for the MAXCUT problem in this subsection in QAOA-like fashion.
Aiming to let DQAS find ansatz without imposing strong QAOA-type assumptions on the circuit architecture, we expand the operation pool with additional Hamiltonians of the form $\hat{H} = -\sum_{\langle ij \rangle} O_iO_j$ and $\hat{H} = \sum_i O_i$, where $O \in \{X,Y,Z\}$; and we refer to these operations as the xx-layer, rx-layer, rz-layer and so on.
In addition, we also add the transversal Hadamard gates and denote it as the H-layer. All these primitive operations can be compiled into digital quantum gates exactly.

Next, let us elaborate on an interesting account that DQAS automatically re-discovers the standard QAOA circuit for the MAXCUT problem.  To begin, we distinguish two settings: instance learning (for a single MAXCUT problem) and ensemble learning (for MAXCUT problems on ensemble of graphs).  As noted in \cite{Brandao2018}, the expected outputs by an ensemble of QAOA circuits (defined by graph instances from, say, Erd\"os-R\'enyi distributions or regular graph distributions) with fixed variational parameters $\theta$ are highly concentrated.  The implication of such concentration is that the optimal parameters (for an arbitrary instance in the ensemble) can be quite close to being optimal for the entire ensemble of graph instances.
This fact not only increases the stability of the learning process with an ensemble of data inputs, but also makes QAOA more practical when the outer optimization loop can be done in this once-for-all fashion. 
In this work, we apply DQAS to both instance learning task and regular graph ensemble learning task \cite{Note1}.

For an ensemble learning on regular graph ensemble (node 8, degree 3), we let DQAS search for an optimal circuit design with $p=5$. By using the aforementioned operation pool comprising the H-layer, rx/y/z-layer and zz-layer with the expected energy for the MAXCUT Hamiltonian as objective function, DQAS recommends the optimal circuit with the following layout: H, zz, rx, zz, rx layers, which coincides exactly with the original QAOA circuit. For metrics in the searching stage, see Fig.~\ref{fig:loss}.
\begin{figure}[t]
	\includegraphics[width=0.35\textwidth]{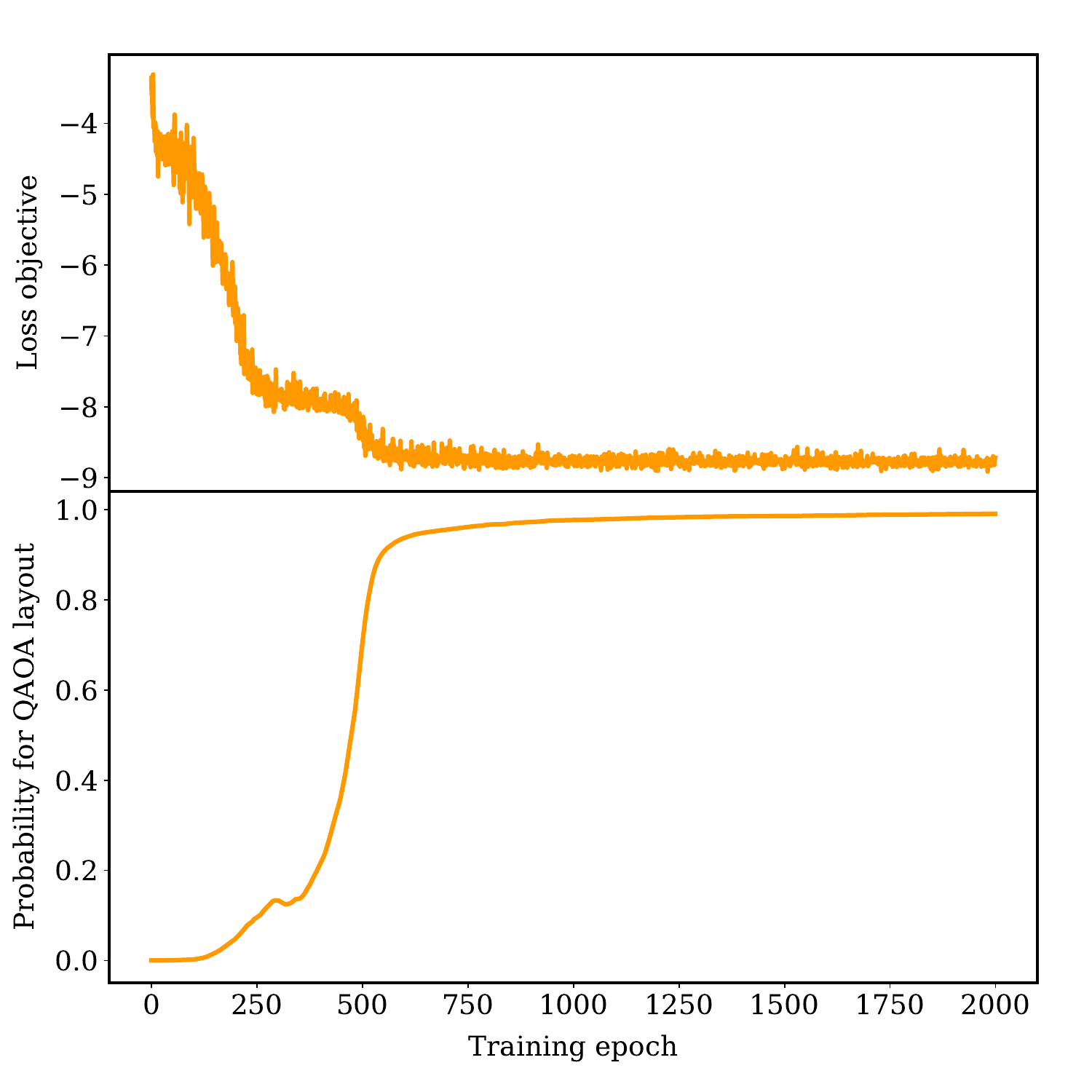}
	\caption{Metrics on DQAS training (depth $p=5$) for MAX CUT problem of degree-3 regular graph ensemble with 8 nodes. The upper plot shows the expected energy/averaged cut value in the training process, the loss is approaching $-8.8$ which reflects the result from $P=2$ QAOA with H, zz, rx, zz, rx layer arrangement. The lower plot indicates how the probability of such optimal layout is increased when the probabilistic model underlying is updated.
	}\label{fig:loss}
\end{figure}

We also carried out DQAS on QAOA ansatz searching with multiple objective consideration on hardware details as well as double-layer block encoding for operations. For details, see the Appendix I \cite{Note1}.

~\newline
\noindent{\bf Reduced graph ansatz searching.} To the end of designing circuits shallower than QAOA, another approach worth attempt is to re-define the primitive circuit layers in the operation pool.
For instance, the zz-layer block is usually generated by the Ising Hamiltonian with the full connectivity of the MAXCUT problem.
However, if the underlying graph of a zz-layer is only a subgraph then the number of gates would be reduced. Suppose we now replace the standard zz-layer (with full connectivity of the original problem) with a set of reduced zz-layers (each generated by a subgraph containing at most half of all edges in the original graph), then a circuit comprising 2 such reduced zz-layers is shallower than the standard $P=1$ QAOA circuit.
As summarized below, ansatz built from such reduced zz-layers is more resource efficient and outperforms the vanilla QAOA layout using the same number of quantum gates. Fig.~\ref{fig:reducedansatz} summarizes the DQAS workflow in searching ansatz with reduced zz-layers.

To demonstrate the effectiveness of this strategy, we consider the circuit design under instance learning setup in which reduced zz-layers in the operation pool are induced by the graph connectivity of a particular instance. In this numerical experiment, we again set out to design a $p=5$ circuit for $n=8$ qubits. More specifically, we generate 10 subgraph with edge density lower than half of the base graph and substitute the base zz-layer with these 10 newly introduced reduced zz-layers in the operation pool.
In such a setup, DQAS is responsible for finding (1) an optimal circuit layout of different types of layers, (2)  best reduced graphs that give rise to the zz-layer in circuit, and (3) optimal parameters $\theta$ for rx/y/z-layer and zz-layer.

Here we give a concrete example. For an arbitrary graph instance drawn from the Erd\"os-R\'enyi distribution with a MAX CUT of 12, DQAS automatically design a circuit that exactly predicts the MAX CUT of 12.  This $p=5$ circuit is composed of following layers: rx-layer, zz-layer, zz-layer, ry-layer and rx-layer. Note the two zz-layers are induced by distinct sets of underlying subgraphs with only four edges each. As a comparison, the $P=1$ vanilla QAOA gives expected MAX CUT of $10.39$, while $P=2$ vanilla QAOA predicts $11.18$. 
In terms of overlap with exact MAX CUT configuration state, the reduced ansatz found by DQAS has nearly $100\%$ success probability for one-shot measurement to get the MAX CUT value while $P=2$ vanilla QAOA has $47\%$ success probability to get the correct MAX CUT value.
The reduced ansatz designed by DQAS consumes about the same amount of quantum resources as the $P=1$ QAOA circuit yet even outperforms the vanilla $P=2$ QAOA circuit. We stress that such an encouraging result is not a special case. By using the reduced ansatz layers, we can consistently find reduced ansatz that outperforms vanilla QAOA of the same depth for MAX CUT problems on a variety of unweighted and weighted graphs \cite{Note1}.

\begin{figure}[t]
	\includegraphics[width=0.47\textwidth]{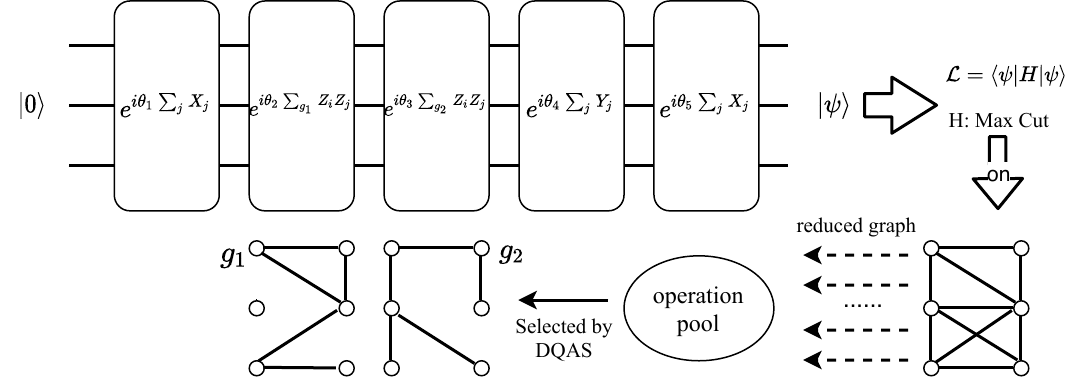}
	\caption{Schematic workflow for reduced graph ansatz search on MAX CUT setup.
		In reduced ansatz searching, there are various reduced graph backend zz-layer in the operation pool.
		These reduced graph are sub graph instances from the problem graph instance. DQAS can not only find the optimal layout and optimal parameters $\vect{\theta}$, but also find the best reduced graph for these zz-layers.
	}\label{fig:reducedansatz}
\end{figure}

DQAS not only can learn QAOA from scratch, but also can easily find better alternatives with shorter circuit depth with an operation pool using slightly tweaked Hamiltonian evolution blocks as primitive circuit layers.  This last achievement is of paramount importance in the NISQ era where circuit depth is a key limitation.

\section{Discussions}

DQAS is a versatile and useful tool in the NISQ era. Not only can DQAS handle the design of a quantum circuit, but it can also be seamlessly tailored for a specific quantum hardware with customized noise model and native gate set in order to get best results for error mitigation.
We have demonstrated the potential of DQAS with the following examples:  circuit design for state preparing and unitary decomposition (compilation), and noiseless and noisy circuit design for the hybrid quantum classical computations.
In particular, we also introduce the reduced ansatz design that proposes shallower circuits that outperforms the conventional QAOA circuits that are inherently more resource intensive ansatz.

In conclusion, we re-formulate the design of quantum circuits and hybrid quantum-classical algorithm as an automated differentiable quantum architecture search. Inspired by DARTS-like setup in NAS, DQAS works in a differentiable search space for quantum circuits.
By tweaking multiple ingredients in DQAS, the framework is highly flexible and versatile. Not only can it be used to design optimal quantum circuits under different scenarios but it also does the job in a highly customized fashion that takes into account of native gate sets, hardware connectivity, and error models for specific quantum hardwares.  The theoretical framework itself offers a fertile ground for further study as it draws advanced concepts and techniques from the newly emerged interface of differential, probabilistic, and quantum  programming paradigms.

~\newline

~\newline
\noindent{\bf Note added}\\ 
\noindent After this work was posted on arXiv, a relevant paper \cite{Du2020a} was also posted. This paper also proposed the idea of using quantum architecture search as a promising strategy for designing hardware-specific and noise-resilient quantum circuits. Conceptually, this work shares some similarities with our work. The approach utilized in their work is of random search and evolutionary nature, where the circuit sampling process stays evenly distributed (i.e. a fixed probabilistic model in our context) while our DARTS-inspired workflow iteratively updates both circuit parameters and the circuit-structure probabilistic model. Together, these two works validate the benefits of using QAS framework to optimizie quantum circuits and should help substantially in establishing quantum advantage in the NISQ era.

~\newline
\noindent{\bf Acknowledgments}\\ 
\noindent SXZ and HY are supported in part by the NSFC under Grant No. 11825404. HY is also supported in part by the MOSTC under Grant Nos. 2016YFA0301001 and 2018YFA0305604, the Strategic Priority Research Program of Chinese Academy of Sciences under Grant No. XDB28000000, Beijing Municipal Science and Technology Commission under Grant No. Z181100004218001, and Beijing
Natural Science Foundation under Grant No. Z180010.


%

\newpage

\begin{widetext}
	\renewcommand{\theequation}{S\arabic{equation}}
	\setcounter{equation}{0}
	\renewcommand{\thefigure}{S\arabic{figure}}
	\setcounter{figure}{0}

	\subsection{Glossary for ingredients of DQAS}
The main components for DQAS are summarized below.	
	\begin{table}[htbp]   
		\begin{tabular}{cc}    \toprule    Glossary &  Explanation \\   
			\midrule  
			circuit encoding & The specific arrangement of circuit ``blanks" to be filled by DQAS. \\   
			operation pool & The set contains all possible parameterized unitaries to construct the circuit. \\  
			probabilistic model  & Circuit candidates are sampled from this model. \\    
			parameter pool  & Circuit parameters organized in the form of (position, operator type) tuple. \\  
			 intermediate representation & The discrete valued vector determining the quantum circuit structure.\\
			\bottomrule  
	\end{tabular} 
			\caption{\label{tab:glossary} The glossary table summarizing DQAS components.}  
 \end{table}
	
	\subsection{Connection to DARTS}
	We illustrate how DQAS is related to DARTS in the neural architecture search.
	In particular, we draw attention to a specific variant of DARTS, the probabilistic neural architecture search \cite{Casale2019} that employs a probabilistic model as the backend NAS.  Both frameworks represent the super network (during an architecture search) in terms of a probabilistic model, and relies on Monte Carlo sampling along with the score-function method to evaluate gradients for structural variables etc.
	
	Different from DARTS, the probabilistic description of the super network is not just an optional approach for avoiding memory-intensive operations to deterministically evaluate the super network but rather an indispensable ingredient of DQAS for circuit design in the NISQ era.
	More precisely, the super network analogy of quantum circuit dictates an implementation of a complex (and potentially non-unitary) operation comprising elementary unitraies $U_j$ present in the operation pool,
	\eq{U=\prod_{i=0}^p \sum_{j}\alpha_{ij}U_j.}{arg2}
	
	This complex operation may be implemented in a quantum circuit via methods like linear combination of unitaries \cite{Childs2012}, which is expensive in the NISQ era.
	The alternative based on the probabilistic model tremendously reduces the near-term implementation challenges by sampling a batch of simpler quantum circuits, each only carrying out an easily implementable unitary transformation. Other than this implementation issue in quantum circuits to motivate a probabilistic solution, DQAS and probabilistic DARTS are highly similar.  The comparison is more succinctly summarized in Fig.~\ref{fig:analogy}.
	
	\begin{figure}[t]
		\includegraphics[width=0.26\textwidth]{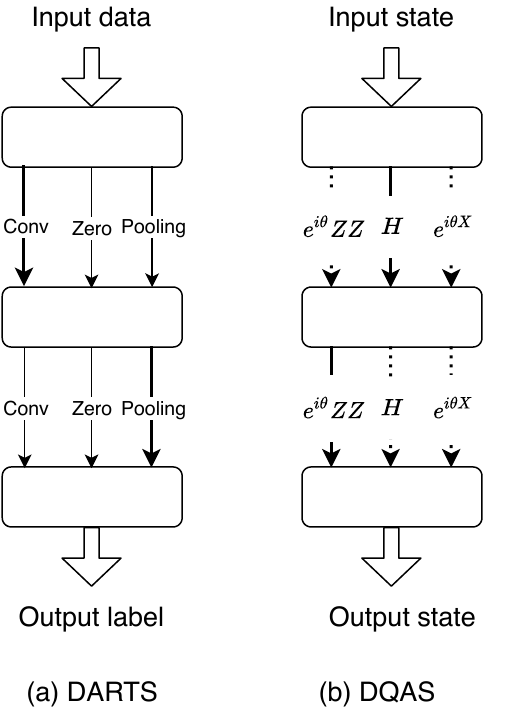}
		\caption{Comparison between setup of (a) DARTS and (b) DQAS in this work. In DARTS, the super network with all paths are evaluated at the same time, while in DQAS, to be in accordance with quantum circuit, only one path is evaluated once as a quantum circuit simulation (indicated by the solid line as shown in (b)), and different path choices are determined by the underlying probabilistic model.
		}\label{fig:analogy}
	\end{figure}
	
	\subsection{Derivation on gradients of probabilistic model parameters}
	
	The forward pass evaluation on the objective reads as:
	\eq{\mathcal{L} = \sum_{\vect{k}\in P} L(\vect{k}) = \sum_{\vect{k}} \frac{P(\vect{k}, \vect{\alpha})}{Z(\vect{\alpha})} L(\vect{k}), }{eq:mcf}
	where we have use $L(\vect{k})$ for the shortcut of $L(U(\vect{k}, \vect{\theta}))$ and we consider unnormalized probability distribution for the general case, where the implicit normalization factor $Z(\vect{\alpha})=\sum_{\vect{k}} P(\vect{k}, \vect{\alpha})$
	
	Directly apply gradient to \Eq{eq:mcf} with score function approach in mind, we have:
	\newcommand{\sxp}{P(\vect{k}, \vect{\alpha})}
	\newcommand{\sxz}{Z(\vect{\alpha})}
	\newcommand{\sxl}{L(\vect{k})}
	\newcommand{\sxn}{\nabla_{\alpha}}
	\newcommand{\sxs}{\sum_{\vect{k}\in P}}
	\al{\nabla_{\alpha}\mathcal{L} &= \sum_{\vect{k}} \nabla_{\alpha} \frac{P(\vect{k}, \vect{\alpha})}{Z(\vect{\alpha})} L(\vect{k})
	\nonumber\\ &=
	\sum_{\vect{k}} \frac{P(\vect{k}, \vect{\alpha})}{Z(\vect{\alpha})} (\frac{\nabla_\alpha{\sxp}}{\sxp}\sxl)-
\frac{\sxp}{\sxz}(\frac{\sxn \sxz}{\sxz}\sxl)\nonumber\\
&=	\sum_{\vect{k}} \frac{P(\vect{k}, \vect{\alpha})}{Z(\vect{\alpha})} (\frac{\nabla_\alpha{\sxp}}{\sxp}\sxl)-(\sum_{\vect{k}} \frac{\sxp}{\sxz}\sxl)(\sum_{\vect{k}}\frac{\sxp}{\sxz}\frac{\sxn \sxp}{\sxp}) \nonumber\\
&=\sxs \sxn \ln \sxp \sxl -\sxs\sxn\ln\sxp \sxs \sxl  .
}
This concludes the derivation of gradient formula for model parameters utilized in the DQAS.
	\subsection{General hyperparameters for DQAS training.}
	We summarize some of the most important ingredients for DQAS below, and leave the extensive investigation on the effects of these adjustable ingredients as well as other ones to future works.
	\begin{enumerate}
		\item Ingredients in common machine learning setup: optimizers, learning rate and schedule for both trainable parameters $\theta$ and structural parameters $\alpha$ of probabilistic models. Since one epoch of evaluation for DQAS is more expensive than conventional neural network evaluations, we may need to find better learning schedules to boost training efficiency for DQAS.
		\item Batch sizes:  This factor plays a very important role in DQAS since score function estimators is in general of high variance. In practice, batch size of $O(100)$ shows good performance in circuit structure searching.
		\item Baselines: there is no theoretical guarantee that running average of objective is the best baseline to lower the variance in Monte Carlo estimations. Therefore, new baselines and even new methods to control variance are worth exploring.
		\item Encoding scheme: as we have seen in examples, different encoding schemes of basic unitary blocks matters in QAS. Therefore, domain specific and expressive encoding scheme beyond simple gate sets, such as the layer and block encoding discussed in the main text, are highly desired for a broader set of applications.
		\item Probabilistic model: The probabilistic model for DQAS can be more sophisticated to better characterize the correlation between layers of circuits. Exploring energy-based models or autoregressive models is a promising future direction for DQAS.
		\item Regularization terms: It is interesting to try and add other regularizations and reward terms into the objectives in order to address multiple objectives such as hardware restriction and quantum noise reduction in the circuit design.
		\item Circuit parameter reusing mechanism: Since the theoretical framework for DQAS is general and can easily go beyond DARTS, we may also explore novel parameter reusing mechanisms beyond the na\"ive ones based on the vanilla super network viewpoint.
	\end{enumerate}
	The following hyperparameter settings are assumed unless explicitly stated otherwise,
	\begin{itemize}
		\item No prethermalization for circuit parameters $\theta$.
		\item Optimizer for probabilistic model parameters $\alpha$ and circuit parameter $\theta$: Adam optimizer with learning step $0.1$.
		\item Initializations: Standard normal distribution for circuit parameters and all zero for probabilistic parameters.
		\item Other techniques: No regularization terms or noise for circuit parameters by default.
	\end{itemize}
	Note that we do not carry out any extensive search for optimal hyperparameter settings. Hence, there is no guarantee that hyperparameters listed below are optimal for corresponding tasks.
	
	For individual tasks and applications in this work, please see the Appendix G for their specific setup and hyperparameter choice.

	\subsection{Training techniques implemented in DQAS}
	There are various training ingredients that can be incorporated into the DQAS framework and many of these tricks and/or improvements are inspired by the works devoted to developing DARTS.  In this subsection, we elaborate on some of these advanced techniques that we have tested.

	~\newline
	\noindent{\bf Multiple starts.} Since the energy landscape for objective functions may be very rugged, parallel training on multiple instances with different fractions of dataset, initialization or randomization scheme may be necessary, where the best candidate circuit with optimal objective value amongst all training instances is returned as the final result.
	
	~\newline
	\noindent{\bf Parameters prethermalization.} Pretraining and updating circuit trainable parameters $\theta$ from the parameter pool for several  epochs as the prethermalization process.  A related topic is parameters initializations.
	Since we would like to search quantum structure without bias, we use all zero initializer for structural probabilistic model parameters $\alpha$.
	For trainable parameters $\theta$, sometimes we can apply domain specific knowledge on the initialization. See QAOA applications in the main text for an example.
	
	~\newline
	\noindent{\bf Early stopping.} To avoid overfitting and reduce runtime, some forms of early stopping may be adopted during the training of DQAS.
	Following the common practices in training DARTS, we may consider typical criteria, such as the standard deviation in a batch of objective evaluations $L$ or the standard deviation in probability of each layer $P$, to decide when to invoke the early stopping.
	The performance gains of DARTS, due to early stopping, were documented in \cite{Liang2019,Zela2019}.

	~\newline
	\noindent{\bf Top-k grid search or beam search.} If the energy landscape is sufficiently complex, DQAS may settle for a local optimal solution instead of a global one.
	In such cases, early stopping can be combined with the so called top-k grid search to avoid trapping into a local minimum.
	Namely, for each layer of the ansatz, we keep the top k (usually $k=2$) most probable operations instead of top-1.
	Therefore, we have $p^k$ candidates for the optimal circuit ansatz. One can easily train these candidate circuits, benchmark their performance , and pick the top performing one as the optimal circuit architecture for a given problem.
	Similarly, we can utilize beam search for optimal structure search, which is the common strategy from text decoder in NLP community. Beam search always maintains $k$ most probable circuit structures and thus avoids the exponential scaling as in grid search.

	~\newline
	\noindent{\bf Baseline for score function estimators.} It is well known that the score function for the Monte-Carlo estimated gradients suffers high variance; although, the situation can somehow be alleviated by the baseline or control variate approach.
	Namely, for the normalized probability distribution $P$,  $\sum_{k\sim P} \nabla \ln P(k) = 0$, one can add any constant in the objectives' gradient as a baseline in order to reduce the variance.
	For instance, during the training of DQAS, we use the running average of $L$ as the baseline. Namely, the loss in Eq.  \eqref{nablaalpha} is actually $L = L-\bar{L}$, where $\bar{L}$ is the average of objective from the last evaluated batch.
	
	~\newline
	\noindent{\bf Layer-by-layer learning.} For deep quantum circuits with large $p$, DQAS may be hard to train from bootstrap.
	Inspired by the progressive training \cite{Chen2019, Skolik2020} for DARTS and quantum neural network training, we apply similar ideas to DQAS.
	Namely, one first find optimal quantum structure with small $p$ and then adaptively increase $p$ by adding more layers to be trained. In this process, one can also reduce the number of candidate operations in the pool based on the knowledge gained from training instances with smaller $p$.
	
	~\newline
	\noindent{\bf Random noise on parameters $\theta$.} The high-dimensional energy landscape can be very rugged in theory. 
	However, based on some numerical evidences, the optimal quantum circuit tends to consistently output similar objective values even when trainable parameters $\theta$ deviate slightly from the optimal values.
	This observation suggests the landscape for optimal quantum structure in terms of trainable parameters is expected to be more flatten than expected. Therefore, to facilitate the search for an optimal circuit architecture, one may add random noise to the trainable parameters $\theta$ to escape the local trapping.
	It is worth noting that the random noises are only added onto trainable/network parameters in our setup instead of structural parameters as in \cite{Chen2020} which tried to bridge the gap of performance between two stages in DARTS.
	
	~\newline
	\noindent{\bf Regularization and penalty terms.}
	Similar to conventional practices in training neural networks, regularization and penalty terms may be introduced in DQAS to avoid overfitting or induce sparsity etc. The applicability of regularization on tunable parameter $\theta$ (with respect to a fixed circuit design) may be easily understood given the high similarity between neural networks and parameterized quantum circuits.
	
	In this section, instead, we focus on the aspect of imposing regularizations on the structural parameters $\alpha$ and manifest the benefits of  regularization on searching for a resource-efficient architecture.
	We provide two concrete examples for illustrations.
	The first example deals with the issue of block merging in DQAS.
	For the simplest probabilistic model for $P(\vec{k},\vec{\alpha})$, each circuit layer is independently sampled. There is a high probability that the same parametrized gates are picked in a consecutive order.
	Namely, the final architecture may contain snippets like $rx(\theta_1)rx(\theta_2)$, which can be easily merged into one layer.
	To address this issue, we propose to add the following terms in the final objective $\mathcal{L}$ to punish such trivial arrangement of circuit layers,
	\eq{\Delta \mathcal{L}_1= \lambda_1\sum_{i=1}^p \sum_{k\in c} p(k_i=k, \alpha) p(k_{i-1}=k, \alpha).}{penalty-merge}
	
	Secondly, since two-qubit gates are primarily responsible for infidelity and errors of quantum computations, it is desirable to select a circuit architecture comprising fewer number of two-qubit gates.  This kind of resource considerations are encouraged during the architecture search if penalty terms of the following form are explicitly added,
	\eq{\Delta \mathcal{L}_2=\lambda_2\sum_{i=1}^p \sum_{k\in c} p(k_i=k)\times\text{\# of two-qubits gates in k}.}{}
	
	We note similar regularization for achieving better training performance \cite{Zela2019} and the multiple objective considerations specifically on computation complexity \cite{Lu2018} have also been reported in the NAS literature.

	~\newline
	\noindent{\bf Proxy tasks and transfer learning.}
	DARTS heavily relies on the idea of proxy tasks to boost performance.
	In DARTS training, one first trains and identifies a suitable network architecutre on the simpler CIFAR-10 (image) dataset; subsequently, one uses the same block topology to build neural networks classifiers for the large-scale ImageNet dataset.
	Same technique may be adapted to DQAS: finding some structure or patterns in quantum circuits for small size problems with small number of qubits or layers and try to apply similar pattern on larger problems.
	For the quantum circuit design, we can even classically simulate the training for small proxy tasks and transfer optimal structures to large problem beyond classical computation power.
	
	We recommend various training techniques, inspired by DARTS-related studies, to obtain more robust and versatile DQAS.
	Due to the close relation between architecture search in the context of quantum circuit and neural network, more interesting ideas may be borrowed from NAS to further improve QAS and innovations in QAS may also inspire developments in NAS.
	
	\subsection{DQAS application in state preparation and circuit compiling}
\noindent{\bf State preparation circuit.}
	We set out to design a quantum circuit for generating GHZ states $\vert \text{GHZ}_n\rangle = \frac{1}{\sqrt{2}}(\vert 0^n\rangle+\vert 1^n\rangle)$ from $\vert 0^n\rangle$.
	To find an optimal structure with less redundancy, we may progressively reduce the layer number $p$ in DQAS until the objective can no longer be accomplished.
	In this case, the operation pool is composed of primitive gates such as single qubit gates and CNOT gates on any pair of qubit.
	In principle, the availability of CNOT gates in the operation pool may be further subjected to the connectivity map of an actual hardware.
	The objective we choose for this problem is the final states distance given by $\sum_i\vert\psi_i-\phi_i\vert$ where $\ket{\phi}$ is the target GHZ state. Such a metric is better to optimize than the typical fidelity or state-overlap objective $\langle \psi\vert \phi\rangle$.
	The optimal circuit found by DQAS is shown in  Fig.~\ref{fig:circuit}\subref{fig:ghz3}.
	It is interesting to observe that DQAS tries to optimize $R_y(\theta)$ by tuning $\theta$ to approximate the behavior of Hadamard gate when Hadamard gate is not given in the operation pool.
	
	\begin{figure}[t]

		\subfloat[]{\label{fig:ghz3}
			\includegraphics[width=0.3\textwidth]{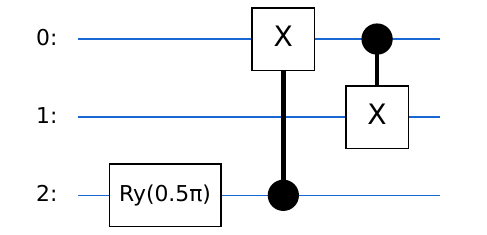}}	
		\subfloat[]{\label{fig:bell}
			\includegraphics[width=0.4\textwidth]{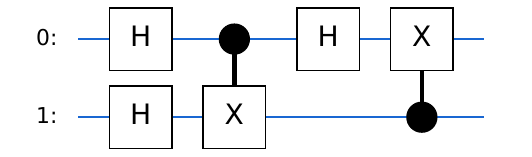}}
		
		\caption{(a) The minimal circuit for preparation of GHZ$_3$ states automatically found by DQAS, where $R_y(\theta) = e^{-i\frac{\theta}{2}}$. ~(b) The circuit with $p=5$ found by DQAS for Bell states transformation.
		}\label{fig:circuit}
	\end{figure}
	
~\newline
\noindent{\bf Unitary decomposition.}
	For state-preparation example, we only care about how the circuit acts on the input state $\vert 0^n\rangle$ and ignore how other input states could be transformed.  This lack of consideration is obvious in the chosen objective above.  For the current example, we aim to decompose arbitrary unitray operation into a set of primitive quantum gates and this implies all inputs transformation are considered.
	For a concrete illustration, we use DQAS to design a quantum circuit for 2-qubits Bell state generation, which is useful for superdense coding.
	We need $2^2=4$ independent input-output pairs to fully characterize the two-qubit unitary under investigation. For instance, the Bell state preparation circuit needs to conform to the following input/output relations, convert the input $\ket{00}$ and $ \ket{11}$ to the Bell states  $\re{\sqrt{2}}(\ket{00}\pm\ket{11})$, and convert the input $\ket{01}$ and $\ket{10}$ to $\re{\sqrt{2}}(\ket{01}\pm\ket{10})$, respectively.
	
	In the next example, we illustrate how to assemble this Bell-state preparation circuit with a finite set of quantum gates.  In other words, we purposely restrict the operation pool to a finite number of discrete gates without any trainable parameters such as rotation angles.  To apply DQAS to search for a Bell-state preparation circuit, we use the input/output relations in Table.~\ref{table:bell} to build the objective function,
	
	\al{\mathcal{L} = -\langle Z_0Z_1\rangle_{U\ket{00}} -\langle X_0X_1\rangle_{U\ket{00}}\nonumber\\
		+\langle Z_0Z_1\rangle_{U\ket{01}} +\langle X_0X_1\rangle_{U\ket{01}}\nonumber\\
		+\langle Z_0Z_1\rangle_{U\ket{10}} -\langle X_0X_1\rangle_{U\ket{10}}\nonumber\\
		-\langle Z_0Z_1\rangle_{U\ket{11}} +\langle X_0X_1\rangle_{U\ket{11}}
		,}
	where $U$ is the tentative circuit proposed by DQAS.
	The final ($p=5$-depth) circuit obtained via DQAS  is presented in Fig.~\ref{fig:circuit}\subref{fig:bell}.
	
	\begin{table}[]
		\begin{tabular}{|c|c|c|c|}
			\hline
			input             & output                                                & $Z_0Z_1$ & $X_0X_1$ \\ \hline
			$\vert 00\rangle$ & $\frac{1}{\sqrt{2}}(\vert 00\rangle+\vert 11\rangle)$ & +1       & +1       \\ \hline
			$\vert 01\rangle$ & $\frac{1}{\sqrt{2}}(\vert 10\rangle-\vert 01\rangle)$ & -1       & -1       \\ \hline
			$\vert 10\rangle$ & $\frac{1}{\sqrt{2}}(\vert 10\rangle+\vert 01\rangle)$ & -1       & +1       \\ \hline
			$\vert 11\rangle$ & $\frac{1}{\sqrt{2}}(\vert 00\rangle-\vert 11\rangle)$ & +1       & -1       \\ \hline
		\end{tabular}
		\caption{Specification of a Bell circuit.}
		\label{table:bell}
	\end{table}
	
~\newline
\noindent\subsection{Hyperparameter settings and training ingredients in experiments}

\noindent{\bf State preparation circuit for GHZ$_3$.} Primitive operation pools: parameterized $R_y$ gate on qubits 0,1,2, CNOT on (0,1); (1,0); (1,2); (2;1) since we consider circuit topology with the nearest neighbor connections. Batch size is 128. Initializer for circuit parameters are zero initializer. Optimizer for $\alpha$: Adam with $0.15$ learning rate.

~\newline	
\noindent{\bf Bell state circuit.} Primitive operator pool: $X, Y, H$ and CNOT gates on each qubit. Note the whole set of operators in the pool is trainable parameter free. Batch size: 128. Optimizer for $\alpha$: Adam with 0.15 learning rate.
	
~\newline
\noindent{\bf QEM on QFT-3 circuit.} Primitive operation pool includes discrete gates: X, Y, Z, S, T, and I (the identity gate). $S = diag(1,i)$ and $T=diag(1, e^{\frac{\pi}{4}i})$.  The I gate must always be in the pool as it stands for leaving the qubits in an idle state. Batch size is 256.
	
	The objective for this QEM task is to maximize the fidelity between noisy output of DQAS-designed quantum circuit and the ideal output.
	In principle , it should be evaluated from a batch of different input states for each circuit.  However, as observed in numerical tests, the standard deviation of the fidelity between noisy and ideal circuit for different input states is small. Therefore, following the spirit of the stochastic gradient descent, the fidelity of such circuit is only evaluated for {\bf one} random input state of each circuit in one epoch. Such a random input state is drawn from the random Haar measure, and this can be partially achieved by a short-depth circuit denoted as the unitary 2-design \cite{Nakata2017}. We utilize 4 blocks repetition of unitary 2-design as input states preparation circuit by default. Since we only want to evaluate the noise in the QFT circuit, we assume the preparation circuit for the random input states noiseless.
	
~\newline
\noindent{\bf QEM on QFT-4 circuit.} For the QFT-4 circuit in the main text, there are 12 slots in total. Therefore, we set $p=12$ for this DQAS design. Since the search space is very large, the search tends to be trapped in a local minimum. Nevertheless, the designed (and potentially sub-optimal) circuits usually outperforms the bare QFT-4 circuit in terms of the fidelity. To reduce the multi-start number, we can restrict the search space by limiting the number of single-qubit gates in the pool. Knowledges on the relevant set of single-qubit gates for such a task can be learned from similar examples such as a QFT-3 case. This procedure utilizing prior knowledge observed in smaller systems and pruning of possible operators follows the philosophy of the progressive training for DARTS \cite{Chen2019} as well as the idea of transfer learning.
	In particular, in this study, the operation pool contains the I placeholder, Z gate, $Z^{\frac{2}{3}}$, $Z^{\frac{4}{3}}$, S gate and T gate.
	Note that there is another trick that may further increase the probability of finding highly nontrivial QEM circuits.
	The idea is to exclude the I placeholder from the operation pool. Without the I placeholder, we force the DQAS engine to fill every gap in the circuit while attempting to maintain a high fidelity. In this way, we encourage DQAS to find nontrivial filling pattern containing long-range correlations of added gates in a QEM circuit and we can recover the nontrivial QEM circuits more easily as shown in the main text. The batch size is $256$. Optimizer for $\alpha$ is Adam with learning rate $0.03$ to $0.06$.
	
~\newline	
\noindent{\bf Typical setups for QAOA ansatz searching.}
	We consider both small and large operation pools for the layer encoding. The small one includes rO-layer with $O=x,y,z$, zz-layer and H-layer. The large one also includes the xx-layer and yy-layer. We have also tested an extra large operation pool with NNN-layers. Namely, we also consider Hamiltonians in the form of ZZ, XX, and YY that couples pairs of next nearest neighbors on the underlying graph. We have reproduce QAOA type layout successfully in such extra large pool. For block encoding scheme, the operator pool includes H-layer, rx-zz-block, zz-ry-block, zz-rx-block, zz-rz-block, xx-rz-block, yy-rx-block, rx-rz-block. For example, zz-ry-block represents for operation on the circuits as $e^{i\theta_m \sum_{\langle ij\rangle}Z_iZ_j}e^{-i\theta_n\sum_i X_i}$.
	
	For the setup of reduced ansatz searching, the operation pool includes H-layer and r-O layers as well as zz-layers but now with different sets of edges.
	We often include 8 to 12 different subgraphs from the problem graph instance.
	And each of them only have a small part of edges as the original graph.
	These subgraphs can be chose randomly and in general $O(10)$ of them is well enough to search for some better ansatz than plain QAOA.
	However, the caveat is that: the number of such reduced graph based zz-layer may have to be larger with graph of more nodes/problems involving more qubits.
	It is more interesting that the reduced graph for these new zz-layers are not necessarily exact subgraph of the problem under investigation.
	These reduced graph can also have random edges and random weights on them, and the result found by DQAS can be as good as subgraph ansatz sometimes.
	The ablation study on reduced graph instances design is an interesting future direction.
	
	We have tried various combinations of ingredient in QAOA ansatz searching. Some are of particular value including: large batch size, typically $64$ to $512$; noise on circuit parameters in simulation, typically independent noise on each parameters in the pool as zero centered Gaussian distribution with standard deviation $0.2$.
	Different objectives, of which CVaR gives promising result apart from conventional energy expectation objectives.
	And when CVaR is concerned, these is no need to add trainable parameter noise in general.
	Initializers for circuit parameters, Gaussian initializer with narrow width and the mean value around $0.2$ to $0.3$ which is near the QAOA optimal parameters region.
	For reduced ansatz search which goes far beyond vanilla QAOA layout, some initializer with larger standard deviation around $1.0$ is utilized.
	SGD optimizer may be better in some cases for updating $\alpha$ with learning rate $0.15$ to $0.3$ and even larger when the gradient is small.
	The learning rate for circuit parameters should be small as $0.005$ to $0.05$ that depends on batch size.
	We also tried L-BGFS and Nelder-Mead optimizers for circuit parameters updates and see no obvious improvements in terms of final objective values.
	Fixing header operator as traversal Hadamard gates boost the training but it is not necessary: the initialization with $\otimes^n H$ can also be auto found via DQAS itself.
	Penalty terms as in the main text with $\lambda_1$ around $0.1$ to $0.2$ may help to alleviate block merging in the training.
	The resource efficiency regularization terms as mentioned in the main text has a smaller $\lambda_2=0.01$. Such term is particularly useful to avoid early attraction by xx-layer and yy-layer (or one can simply drop xx-layer and yy-layer from the beginning as they are shown to be redundancy in the main text).
	
	Other techniques include those discussed in the main text: top-k grid search postprocessing and multi start may be necessary to find optimal structure as the energy landscape of such search is rather complicated and lots of lower p QAOA layouts serve as local minimum traps.
	For example, if we carry out DQAS for 5 layers, we will often end in an architecture equivalent to $P=1$ QAOA instead of $P=2$.
	The reason behind that is the performance improvement with deeper QAOA layout is slight and the global minima is vagued by lots of local minimum with similar objective values.
	Block encoding with two layers combination as primitive operators in the pool is much better to train than layer encoding scheme and mitigate the above problem. On the contrary, novel objectives such as Gibbs objective that shows sharper energy landscape in circuit parameter space are not suitable in DQAS when bi-optimization dominates and flatten energy landscape helps.
	It is also worth mentioning that, in ensemble learning setup, the dataset/graph instances set is not pre-determined. Instead the graph instance is generated on the fly in DQAS training and in princinple one can experience all graph instances of one ensemble as long as the searching epochs is large enough. This design is better than training on fixed dataset of ensemble of graph instances, which tends to over fitting.
	
	Construction of QAOA primitive layers with native quantum gates set as follows:
	The single-qubit layer specified in the form of $e^{-i \theta\sum_i O_i }$ is just a single-qubit rotation gate $r_O$.
	The layer of two-qubit gates is of the form  $e^{i \theta\sum_{\langle ij\rangle} O_i O_j}=\prod_{\langle ij\rangle}e^{i\theta O_iO_j}$, and each term $e^{i \theta O_iO_j }$ can be implemented as in Fig.~\ref{fig:zz}.
	If $O$ is not Z, then basis-rotation gates (Hardmard $H$ for $O=x$, $R_x(\pm \frac{\pi}{4})$ for $O=Y$) are attached on both sides of the circuit.
	From this perspective, xx-layer and yy-layer are actually redundant since they can be exactly implemented by using H-layer+zz-layer+H-layer or rx-layer+zz-layer+rx-layer, respectively.  Therefore we can safely drop the xx-layer and yy-layer from the operation pool.

	\begin{figure}[t]
		\includegraphics[width=0.3\textwidth]{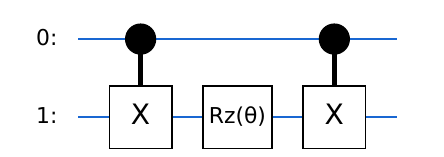}
		\caption{The circuit construction for $e^{-\frac{i\theta}{2}Z_0Z_1}$ by CNOT and $R_z$ gate. For implementation of $e^{-\frac{i\theta}{2}X_0X_1}$ and $e^{-\frac{i\theta}{2}Y_0Y_1}$, the only change is to pretend and append Hadamard gate or $R_x(\pm \frac{\pi}{4})$  gate on both sides and each qubit lines. This is the key building block for QAOA layers.
		}\label{fig:zz}
	\end{figure}

\subsection{Further results on QEM of QFT circuit}
\noindent{\bf Fidelity results for bare circuit, na\"ive QEM circuit and nontrivial QEM circuit by DQAS.}
	The noise model for the backend quantum simulation is described in the main text. The bit-flip errors are randomly inserted between adjacent circuit layers. Whenever there is an empty slot (idel state) in the circuit, the error rate is larger.  Next, we not only present a comparative study between QEM circuits found by DQAS and bare circuit, but also present a comparative study of QEM circuit designed by DQAS and QEM circuit inspired by  theoretical methods.
	
	For QFT-3 circuit,  the typical textbook circuit only gives a fidelity of $0.33$ in the presence of the bit-flip errors;
	naive QEM circuit with additions of pairs of Pauli gates on qubit 0 and 2 gives an ameliorated fidelity of $0.55$ (different choices of Pauli gates only yield a small differences in the fidelity). QEM circuit found by DQAS gives a further improved fidelity of $0.6$.
	
	For QFT-4 circuit, the typical textbook circuit gives a fidelity of $0.13$. Again, the na\"ive QEM circuit, where as many pairs of Pauli gates as possible are added to fill empty slots (see Fig.~\ref{fig:policy}\subref{fig:naive}), gives a fidelity of $0.3$. There is another type of circuit-filling policy (see Fig.~\ref{fig:policy}\subref{fig:advance}), where we do nothing to single empty slots in a circuit, and insert a series of $Z^{\alpha_i}$ that collectively give an identity into contiguous empty slots (spanning across more than 2 layers). This strategy allows one to fill empty slots in the circuit except those isolated ones restricted to a single circuit layer. QEM circuit designed under this policy gives a fidelity around $0.41$. On the other hand, DQAS discovers many distinct configurations of QEM circuit for the QFT-4 case, and the associated fidelities are usually found to be in the range of $0.45\sim 0.46$. Clearly, automated circuit designs outperform the ones recommended by human-designed and sophisticated empty-slot-filling policy .
	
	\begin{figure}[t]

		\subfloat[]{\label{fig:naive}
			\includegraphics[width=0.38\textwidth]{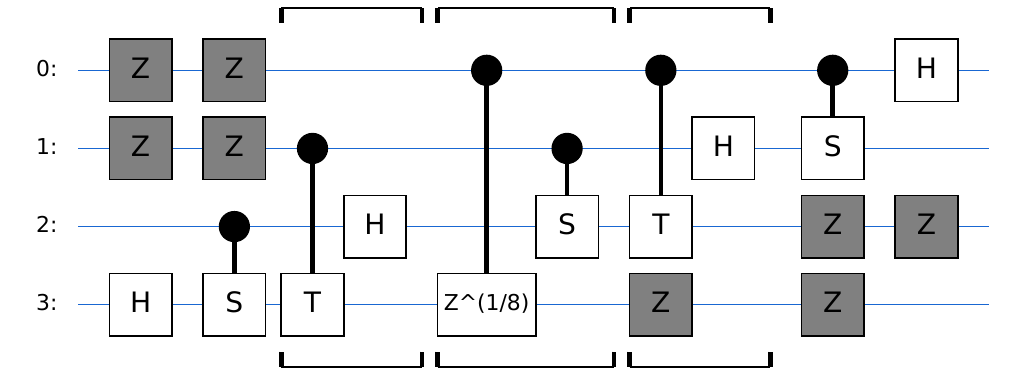}}
		\subfloat[]{\label{fig:advance}
			\includegraphics[width=0.45\textwidth]{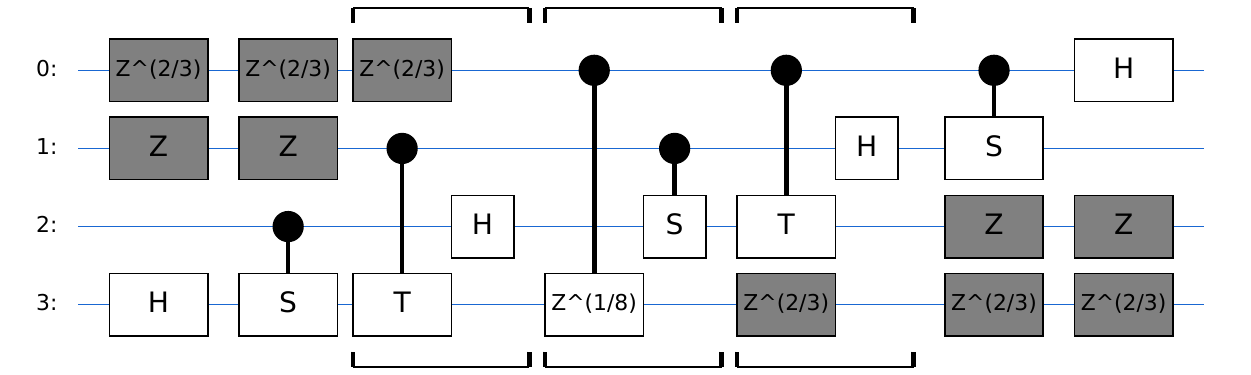}}
		
		\caption{Some human design gate inserting policies for QEM on QFT-4 circuit. (a) The na\"ive Pauli pair inserting whenever possble. (b) The advance inserting which tries to collectively return identity in each gap except from single holes.
			DQAS found QEM circuit can outperform these circuits in terms of fidelity.
		}\label{fig:policy}
	\end{figure}
	
~\newline
\noindent{\bf More QEM circuits with similar fidelity for QFT-4.} 
	In QFT-4 case, DQAS also finds various circuits of similar fidelity as the optimal one in the main text, we present some of them in Fig.~\ref{fig:moreqem}.
	
	\begin{figure}[t]

		\subfloat[]{
			\includegraphics[width=0.43\textwidth]{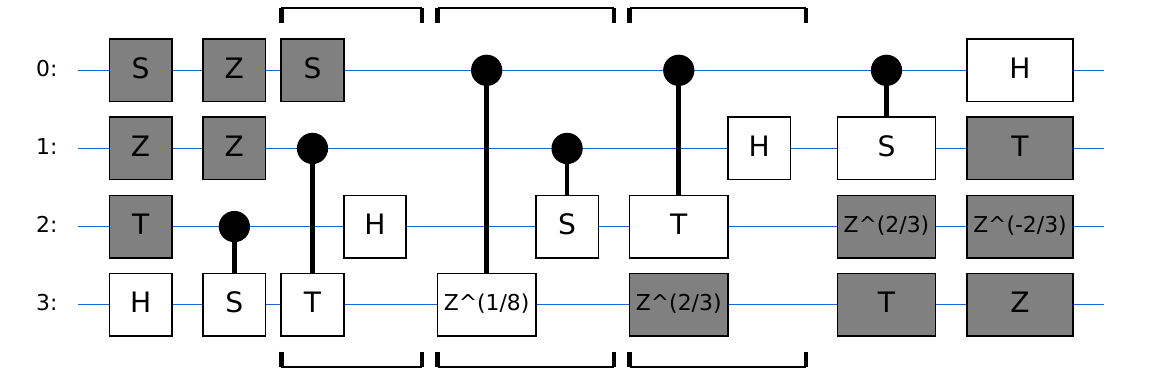}}
		\subfloat[]{
			\includegraphics[width=0.4\textwidth]{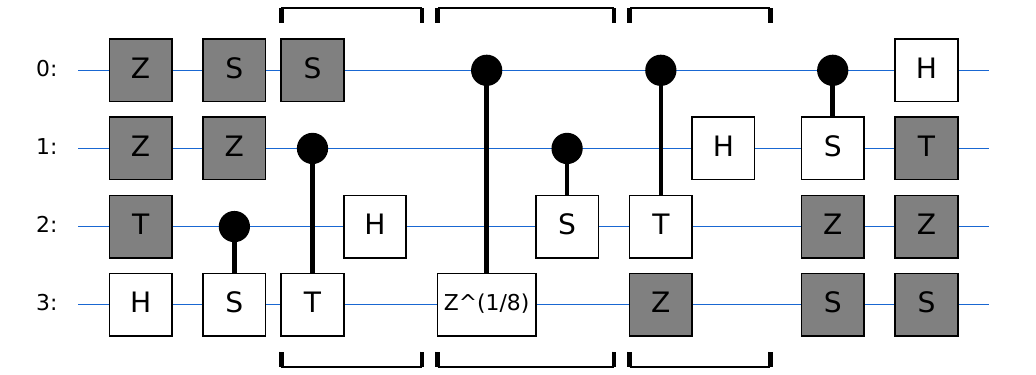}}
		
		\caption{Some optimal QEM circuit found by DQAS. They shows similar performance in fidelity as the one given in the main text.
		}\label{fig:moreqem}
	\end{figure}
	
\subsection{Further results on QAOA ansatz searching}
\noindent{\bf Multiple objectives with hardware consideration.}
	Additional considerations may be taken into account during the search for an ideal circuit design for the MAXCUT problem.  Suppose that we still work with the layer-encoding operation pool given above but with xx-layer, yy-layer explicitly considered. Furthermore, we suppose the backend quantum hardware  is equipped with the primitive gates of rO, H, and CNOT.  Therefore, every circuit layer has to be translated into this native gate set. To design resource-efficient quantum circuit by DQAS, we may add the following penalty term to incorporate consideration of resource limitation and quantum error mitigation into the mix,
	\eq{\nabla \mathcal{L} =\lambda_2 \sum_{i=1}^p \sum_{k\in c} p(k_i=k) \omega(k),}{qaoa-penalty}
	where $\omega(\text{CNOT}) = 2$, $\omega(\text{rO}) = 1$ and $\omega(\text{H}) = 1$.  Given the above costs for each gate, one can easily show that $\omega(\text{xx-layer}) = 27/2$ for regular graph of degree 3.
	These weights can be further adjusted as hyperparameters to reach better multiple objective frontier on both performance and resource efficiency.  Within such a large operation pool and regularization terms as Eq. \eqref{qaoa-penalty}, we can again resume typical QAOA layout as H, zz, rx, zz, rx.
	Furthermore, we can find other competitive structures beyond the vanilla QAOA layout as shown in the following information.
	
	~\newline	
	\noindent{\bf Block encoding for QAOA search.} When dealing with large-scale circuits, it may become progressively more challenging to discover an optimal structure for MAX CUT problems with DQAS.
	This is partially related to our decision to utilize the simplest probabilistic model, where no explicit correlations about choices of consecutive layers are taken into account of. It might be useful to add these correlations into the design.  For instance, in the standard QAOA circuit, an rx-layer is always followed by a zz-layer.
	
	Acknowledging the usefulness of such a complex block that contains at least two primitive layers, we introduce the block encoding. Namely, the primitives in the operation pool are of the form $e^{i\theta_1 ZZ}e^{-i\theta_2 X}$.  In other words, the block encoding deals with various combinations of basic operations like zz-rx-block, yy-rz-block and so on. Same as before, it is useful to keep the Hadamard H-layer in the pool.
	Via this block encoding, we easily discover the standard QAOA layout for $P=3$ as H-layer, zz-rx-block$* 3$ in 8 qubit system. The emergence of a clear pattern of circuit layout indicates that zz-x-block holds the key for making highly accurate predictions on the MAXCUT problems.  Thus, DQAS learns the essence of QAOA circuit purely by exploring and assessing different possibilities without imposing strict assumptions.

	~\newline
	\noindent{\bf Proxy tasks and transfer learning.} We mainly apply DQAS to design QAOA-like ansatz for systems consisting of $8$ or $10$ qubits.
	We can think of these automated designs, conducted for small systems, as proxy tasks since large systems (containming more qubits) often share the same optimal circuit patterns with small systems of the same family of problem like MAXCUT. Then we can fix the circuit architecture and only optimize circuit parameters for larger systems as in the standard QAOA algorithm.
	
	~\newline
	\noindent{\bf Other competitive and near optimal layouts found by DQAS.}
	In the study of a bunch of random regular graphs,  DQAS not only finds QAOA circuits but also discovers other circuit designs with comparable or slightly inferior performance. More specifically, via an ensemble made of $n=8$ regular graphs with degree $3$, we found an alternative architecture comprising of H-layer, yy-layer, rx-layer, zz-layer, rx-layer, which gives an expectation value of MAX CUT around $8.8$ which is similar to the performance of QAOA layout of the same depth.
	
	~\newline
	\noindent{\bf Layerwise training.} It becomes increasingly challenging to find an optimal ansatz for deeper circuit, since there are multiple local minima on this rugged energy landscape for MAXCUT problems. One way to understand the challenge is to note that all QAOA circuits that effectively are constituted of $P-1$ blocks are local minima for $P$ blocks QAOA.
	When the $P$-block optimal circuit delivers a marginal performance gain with respect to the $(P-1)$-block optimal circuits, the difficulty of doing a brute force search  for an optimal circuit layout increases with DQAS depth $p$. To resolve this challenge, the idea of layer-wise progressive training should be adopted \cite{Skolik2020}.
	Note that multiple-start technique is necessary due to the hardness of finding optimal circuits when the circuit depth is large. Early stopping as well as top-2 grid search are also found to be useful.
	
	~\newline
	\noindent{\bf  Erd\"os-R\'enyi graph results.} In this study, we also try out DQAS design for MAXCUT problems with the graphs drawn from the  Erd\"os - R\'enyi ensemble.  Moe specifically, we consider two ensembles characterized, respectively, with $n=10, p=0.3$ and $n=8, p=0.4$ .  Interestingly, DQAS rediscover QAOA layout as an optimal architecture under the setting of ensemble learning in this case too.  Note that an ensemble of Erd\"os - R\'enyi graph usually suffers greater variance, in terms of the exact MAX CUT value and the optimal circuit parameters across each graph instance, than that of a regular-graph ensemble.
	
	~\newline
	\noindent{\bf Different objectives beyond an average value.} Since the aim of QAOA is to estimate the lowest energy instead of an average one,  it seems to be natural to use objectives that may more accurately reflect the overlap between  the prepared quantum state and the ground state instead of the average energy. To this end, there are some newly proposed objectives such as the CVaR \cite{Barkoutsos2020} and Gibbs objective \cite{Li2020}. These objectives can be easily handled within the framework of DQAS with customized objectives.  For example, for Gibbs objective, we simply set $f(x)=e^{-\lambda x}$ and $g(x) = \ln x$. Gibbs objective is supposed to reward the prepared quantum state to have a higher overlap with lower energy states. However, Gibbs objective is found to introduce a very steep landscape with respect to circuit parameters. Gibbs objective may be useful when the task only requires to search for optimal circuit parameters  for a given circuit architecture.  However, in a typical DQAS-task, we need to identify an optimal circuit design (i.e. simultaneously finding an architecture and related circuit parameters), the sharp landscape of Gibbs objective presents a non-trivial challenge.
	In particular, the circuit parameters in DQAS-designed circuits frequently deviate from the optimal circuit parameters in the searching process due to the complications of the existence of a super network structure, the minimum of Gibbs objective is hence vague.
	
	On the contrary, CVaR seems to be a good objective to try with DQAS. CVaR actually measures the mean energy of only a proportion of samples (say 20\%) having the lowest energies. This objective gives a much smoother energy landscape than that of Gibbs objective by nature.
	
	In short, Gibbs objective is not compatible with the DQAS framework, but CVaR objective shows promising potential in our study.
	The reduced ansatz search on weighted graphs, mentioned in the main text and further described in the next section of this supporting information, is actually based on CVaR objective. We leave detailed comparison and ablation study on different choices of objectives as a future work.
	
	~\newline
	\noindent{\bf Instance learnings.} As we have discussed in the main text, we may design  an optimal circuit for individual problems (dubbed as the instance learning) or an ensemble of problems. It is quite obvious that highly customized circuit architecture, adapting specifics of a particular graph instance, may outperform a generic QAOA layout when the circuit depth is restricted. In this section, we report our study on using DQAS to design ideal circuits for individual instances of the MAXCUT problem.
	
	For example, DQAS recommends an optimal circuit composed of  yy-layer, zz-layer and yy-layer that gives an expected energy of $-8.0$, whereas the vanilla QAOA circuit of the same depth only gives an average energy of $-7.75$. Therefore, such an example demonstrates the effectiveness of using DQAS to design customized circuits in an end-to-end fashion. By applying DQAS, we can both discover a universally optimal architecture for problems under the setting of the ensemble learning, and recommends customized optimal designs for specific problems under the setting of the instance learning.
	
	Different graph instances tend to present distinct searching difficulty for DQAS.  For some graph instance, the performance of an optimal quantum circuit is way better than that of other candidates. There is a significant ``performance gap" among circuit designs; therefore, DQAS can easily identify the optimal circuit in these cases. However, there exists other graph instances in which the performances are rather similar. These scenarios are usually accompanied with a training landscape having many local minima. DQAS may easily get trapped into one local minimum and recommends a sub-optimal circuit in these cases.
	Nonetheless, this is totally acceptable for problems like MAXCUT. For more challenging problem, such as the quantum simulations where the exact ground state is strictly desired, we anticipate additional tweaks to be implemented to enhance DQAS. For instance, one may adopt the idea of an imaginary-time evolution (or natural gradient descent) for the training procedure.
	
	\begin{figure}[t]
		\includegraphics[width=0.28\textwidth]{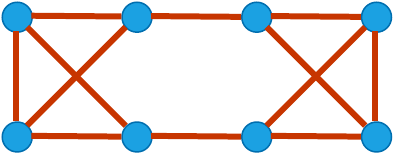}
		\caption{One of the instance learning unweighted graph for MAX CUT problem. The MAX CUT value for this graph is $10$. We can found better quantum architecture than QAOA of the same depth to approximate MAX CUT value in such instances.
		}\label{fig:graph8b}
	\end{figure}
	
~\newline
\noindent{\bf Reduced graph ansatz searching on instance learning.}
	Finally, we elucidate the study of ansatz searching with reduced graphs introduced in the main text.  We provide two examples.
	The first task is to design optimal ansatz circuit for an unweighted graph (shown in Fig.~\ref{fig:unweightedreduce}), and adopt the mean energy of the proposed circuits as the training objective. An optimal circuit, found by DQAS, for this specific graph is rx-layer, zz-layer, zz-layer, ry-layer and rx-layer. Recall the zz-layers are generated with Hamiltonian with restricted connectivity sampled from the original graph as explained in the main text. Figure \ref{fig:unweightedreduce} also gives the two subgraphs used to generate the 2nd and 3rd layer of ZZ Hamiltonians in this optimal design. Finally, we quote the MAX CUT estimated by the optimal circuit and the vanilla QAOA for comparison. For this specific graph, the exact MAXCUT is 12.  The optimal reduced ansatz by DQAS and the $P=1,2$ vanilla QAOA estimate the MAXCUT to be 12, 10.39, 11.18, respectively. In terms of overlap with correct MAX CUT ground state, optimal reduced ansatz has nearly identity overlap while $P=1,2$ QAOA ansatz has overlap $0.250, 0.471$ in value. Note the edge number in these reduced graph is way less than one vanilla zz-layer, the obtained structure is actually much shallower than $P=1$ vanilla QAOA while the performance is better than deeper QAOA layouts.
	
	The second example is to design circuits for a weighted graph with weights distributed as Gaussian distribution $\mathcal{N}(1, 0.2)$. For this case, we adopt the CVaR objective. This specific graph instance, shown in Fig.~\ref{fig:weightedreduce}, has 8 nodes.  The optimal reduced ansatz recommended by DQAS is ry-layer, rx-layer, zz-layer, zz-layer, rx-layer. Again, the two zz-layers are induced by two subgraphs  shown in Fig.~\ref{fig:weightedreduce}. Note the subgraphs inherit the exact weight value of corresponding edges. For this graph, the CVaR results for the exact evaluation, optimal reduced ansatz and $P=1,2$ vanilla QAOA circuits are $10.20587$, $10.18$, $9.63$, and $10.20587$, respectively.
	For this instance, $P=2$ QAOA circuit essentially gives the exact result while the ``optimal" DQAS circuit does not. However, we remind readers that this ``optimal" DQAS circuit is only about the same circuit depth as the $P=1$ QAOA circuit.

	We remark that the reduced subgraphs, in both examples above, have edge density far below half of the base graphs. Therefore,  DQAS can identify suitable circuit to give highly accurate estimations of MAX CUT based on very spare graph connectivity for zz-layers in these illustrations.
	
	There are also some arguments or physical intuitions behind such reduced graph ansatz. It is reminiscent to the random attention or the global attention mechanism \cite{Beltagy2020, Zaheer2020} recently developed for transformer models for the natural language processing. Even the number of edge is way  less than base graph, each node can still be visited via random walk as long as there are enough layers.
	
	\begin{figure}[t]
		\includegraphics[width=0.6\textwidth]{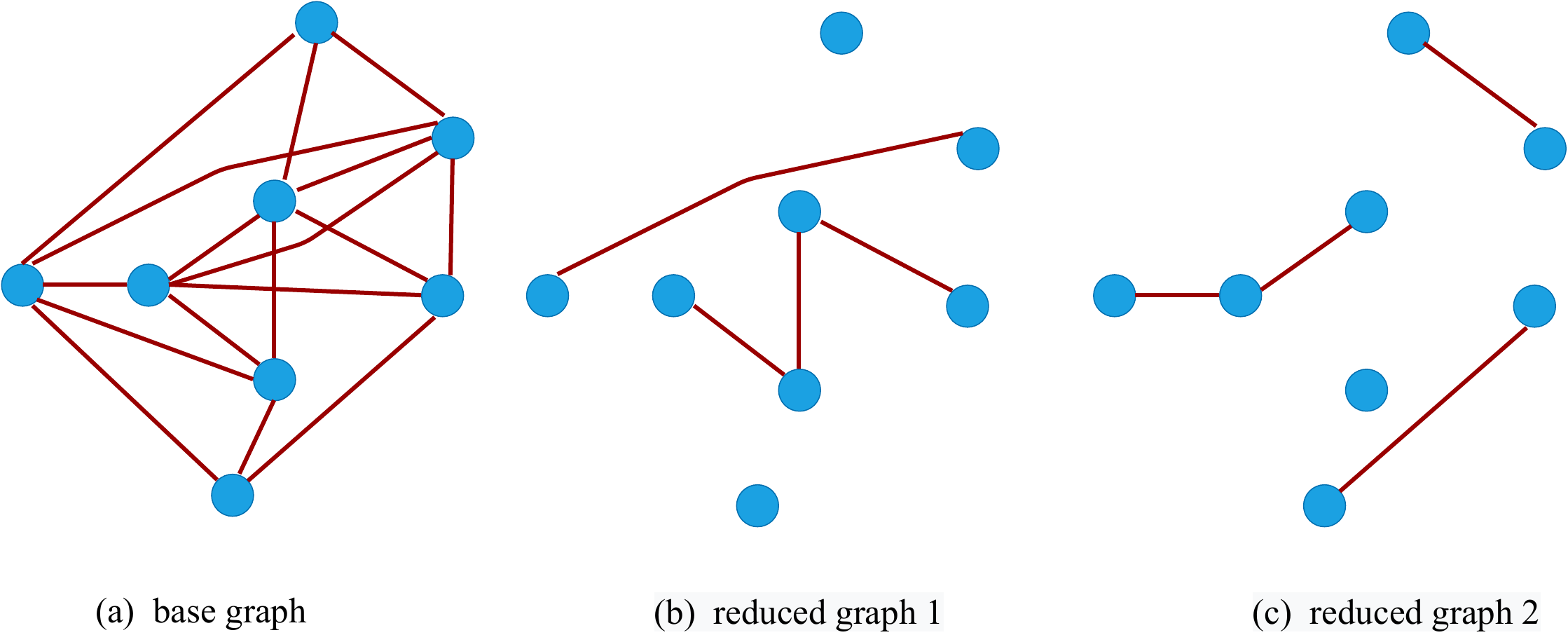}
		\caption{Reduced graph ansatz for unweighted graph case (all weights are unity). (a) Base graph for MAX CUT and (b)(c) reduced graph found by DQAS in reduced ansatz layout which outperforms $P=2$ plain QAOA.
		}\label{fig:unweightedreduce}
	\end{figure}
	
	\begin{figure}[t]
		\includegraphics[width=0.54\textwidth]{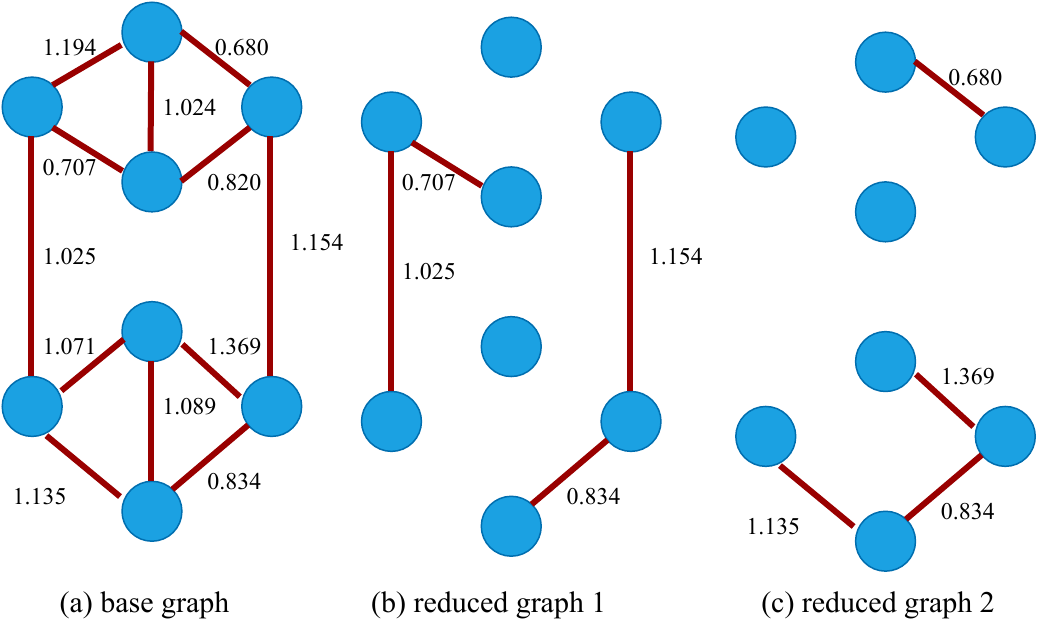}
		\caption{Reduced graph ansatz for weighted graph case. (a) Base graph for MAX CUT and (b)(c) reduced graph found by DQAS in reduced ansatz layout which outperforms $P=1$ plain QAOA.
		}\label{fig:weightedreduce}
	\end{figure}
	
\end{widetext}

\end{document}